\documentclass[fleqn,usenatbib,useAMS]{mnras}

\usepackage{graphicx}	
\usepackage{amsmath}	
\usepackage{amssymb}	
\usepackage{multicol}        
\usepackage{bm}		
\usepackage{pdflscape}	

\usepackage[T1]{fontenc}
\usepackage{ae,aecompl}

\usepackage{newtxtext,newtxmath}

\newcommand\mpo{\textcolor{black}}
\date{Last updated 2020 June 10; in original form 2013 September 5}

\pubyear{2023}

\title[Mixing in core-collapse supernova remnants]{Mixing of materials in magnetised core-collapse supernova remnants}

\author[D. M.-A.~Meyer et al.]
       {D. M.-A.~Meyer\thanks{E-mail: dmameyer.astro@gmail.com}$^{1}$, M.~Pohl$^{1,2}$, 
       M.~Petrov$^{3}$ and K.~Egberts$^{1}$ \\
       $^{1}$ Universit\" at Potsdam, Institut f\" ur Physik und Astronomie, 
              Karl-Liebknecht-Strasse 24/25, 14476 Potsdam, Germany \\
       $^{2}$ Deutsches Elektronen-Synchrotron (DESY), Platanenallee 6, 15738 Zeuthen, Germany \\
       $^{3}$ Max Planck Computing and Data Facility (MPCDF), Gießenbachstrasse 2, D-85748 Garching, Germany \\ 
       }

\begin{document}
\label{firstpage}
\pagerange{\pageref{firstpage}--\pageref{lastpage}}
\maketitle

\begin{abstract}
Core-collapse supernova remnants are structures of the interstellar medium (ISM) 
left behind the explosive death of most massive stars ($\lesssim 40\, \rm M_{\odot}$). 
Since they result in the expansion of the supernova shock wave into the gaseous 
environment shaped by the star's wind history, their morphology constitutes an insight into 
the past evolution of their progenitor star. Particularly, fast-moving massive 
stars can produce asymmetric core-collapse supernova remnants. 
We investigate the mixing of materials in core-collapse supernova 
remnants generated by a moving massive $35\, \rm M_{\odot}$ star, 
in a magnetised ISM. Stellar rotation and the wind magnetic field are 
time-dependently included into the models which follow the entire 
evolution of the stellar surroundings from the zero-age main-sequence 
to $80\, \rm kyr$ after the supernova explosion. 
It is found that very little main-sequence material is present in remnants from moving 
stars, that the Wolf-Rayet wind mixes very efficiently within the $10\, \rm kyr$ 
after the explosion, while the red supergiant material is still unmixed by $30\%$ 
within $50\, \rm kyr$ after the supernova. 
\textcolor{black}{
Our results indicate that the faster the stellar motion, the more complex the 
internal organisation of the supernova remnant and the more effective the mixing of 
ejecta therein. \mpo{In contrast, the mixing of stellar wind material is only weakly affected by 
progenitor motion, if at all.}
}
\end{abstract}



\section{Introduction}
\label{sect:intro}

High-mass stars are defined as stellar objects of mass $M_{\star}$ heavier 
than $8\, \rm M_{\odot}$~\citep{salpeter_apj_121_1955,kroupa_mnras_322_2001}, 
and they represent $\approx 1\%$ of all stars in the Milky Way. 
They form in dense and highly opaque giant molecular clouds, as a result of the 
local free-fall gravitational collapse of dense cold gas~\citep{shu_pavi_book_1992}. 
They follow a formation scenario that is a scaled-up version of that of 
low-mass stars~\citep{voroboyov_apj_650_2006,vorobyov_apj_704_2009,vorobyov_apj_723_2010}, 
and which is well-described by the so-called burst mode of accretion in  
star formation~\citep{meyer_mnras_464_2017,meyer_aa_mnras_2021}. 
When the infalling molecular material lands onto a centrifugally-balanced accretion 
disc, it undergoes gravitational instability and gaseous clumps can form and 
inward-migrate onto stellar surface, adding mass to the young massive 
star~\citep{meyer_mnras_482_2019}. 
Once massive protostars are hot enough, their ionizing feedback dissipate 
their surrounding molecular environment and they release strong supersonic 
winds from the stellar surface~\citep{vink_aa_362_2000} to interact 
directly with the interstellar medium (ISM), giving birth to stellar wind 
bubbles~\citep{weaver_apj_218_1977}.

As a function of multiplicity~\citep{sana_sci_337_2012}, 
metallicity~\citep{brott_aa_530_2011a,brott_aa_530_2011b} and rotation 
properties~\citep{maeder_araa_38_2000,szecsi_aa_658_2022}, massive stars 
evolve from a long initial main-sequence phase by undergoing several 
evolutionary phases, characterised by abrupt surface properties changes, 
alternating between hot, fast-winded with cold, slow-winded 
phases~\citep{langer_araa_50_2012}. 
Although exotic evolutionary phases exist~\citep{smith_araa_52_2014,groh_aa564_2014}, 
a high-mass star typically goes through a red supergiant 
phase~\citep{levesque_newar_54_2010,davis_rspta_2017},  
followed, in the case $M_{\star}\ge 30\, \rm M_{\odot}$, by a Wolf-Rayet 
phase~\citep{underhill_ararr_6_1968,hamman_lnp_401_1992,crowther_araa_45_2007}, 
before exploding as a supernova~\citep{woosley_araa_24_1986,
weiler_araa_25_1988,smartt_araa_47_2009}. 
%
%
%
The circumstellar medium of massive stars is the region of the ISM directly 
surrounding high-mass stellar objects and it adopts morphologies and emission 
properties reflecting both the stellar feedback and the ambient medium properties.

The surroundings of young massive stellar objects are massive accretion discs of 
radius $\sim 1000\, \rm au$~\citep{johnston_apj_813_2015,ilee_mnras_462_2016,  
ahmadi_aa_618_2018} that are evaporated and blown by the radiation 
field and the wind of zero-age main-sequence high-mass stars. 
Their environments adopt the theoretical shape of the so-called stellar wind bubble 
described in~\citet{weaver_apj_218_1977}, which is distorted into a stellar 
wind bow shock~\citep{gull_apj_230_1979,wilkin_459_apj_1996}  
in the case of the $30\%$ of runaway massive stars which move supersonically 
through the ISM~\citep{blaauw_bain_15_1961,gies_apjs_64_1987,gvaramadze_424_mnras_2012, 
Moffat1998}. Such bow shock can be observed throughout the whole electromagnetic 
spectrum, from the ultraviolet~\citep{kaper_apj_475_1997} and the infrared 
waveband~\citep{peri_aa_538_2012,peri_aa_578_2015,kobulnicky_apjs_227_2016, 
kobulnicky_aj_154_2017} to the high-energies~\citep{benaglia_aa_517_2010,   
delvalle_aa_550_2013,deBecker_mnras_471_2017}. 
When the star evolves to the red supergiant phase, the wind interacts either 
with the interior of the main-sequence wind 
bubble~\citep{garciasegura_1996_aa_316ff,garciasegura_1996_aa_305f,
freyer_apj_638_2006}, or directly with the ISM if the star moves fast 
enough~\citep{brighenti_mnras_277_1995,chita_aa_488_2008,herbst_ssrv_218_2022}. 
The circumstellar medium of Wolf-Rayet stars takes the form of circular rings 
in the context of fast-moving, high-latitude objects, bipolar nebulae in the case 
of rotating massive stars or more complex 
morphologies~\citep{brighenti_mnras_273_1995,meyer_mnras_496_2020,meyer_mnras_507_2021}. 
The surroundings of evolved massive stars is therefore the fingerprint of its past 
life onto the ISM, and, it constitutes the location of the supernova explosion. 
The supernova remnants left behind the death of high-mass stars are shaped by the 
interaction of the supernova shock wave with the circumstellar 
medium~\citep{weiler_araa_25_1988}. Unfolding them permit to reconstruct the 
past evolution and to constrain the properties of its progenitor star.

Supernova remnants of massive progenitor stars, or core-collapse supernova 
remnants, are characterised by the presence of circumstellar material enriched 
in the heavy elements synthetised by the progenitor released as stellar winds.  
They reveal the presence of elements such as 
Fe~\citep{uchida_pasj_61_2009}, that have been produced in the core 
of the progenitor and by the possible presence of a neutron star left inside of the supernova 
remnant~\citep{zavlin_aa_331_1998,gvaramadze_aa_454_2006}, ultimately leading to the 
formation of pulsar wind nebulae~\citep{2023arXiv230112903O,olmi_mnras_490_2019}. 
The supernova explosion is closely connected to non-thermal 
physical mechanisms such as the anisotropic liberation of 
neutrinos~\citep{kifonidis_aa_453_2006,mueller_aa_573_2012,mueller_mnras_448_2015} 
and the production of gravitational waves~\citep{jardine_mnras_510_2022}. 
Models from the early instants~\citep{aloy_mnras_500_2021} up to the 
first year of core-collapse event has been subject to numerical modelling 
in~\citet{gabler_mnras_502_2021}. 
The interaction of the liberated supernova shock wave with the last wind of the 
progenitor generates characteristic lightcurve functions of the wind properties,
which have been observed and 
modelled~\citep{chevalier_apj_258_1982,chevalier_apj_344_1989,truelove_apjs_120_1999}. 
The propagating shock wave interacts with the circumstellar material before 
further expanding into the interstellar medium~\citep{tutone_aa_524_2020}, 
leading to the production of non-thermal 
emission~\citep{orlando_aa_622_2019,das_aa_661_2022}.
A well-documented example of such 
mechanism is that of Cassiopeia A~\citep{finn_apj_830_2016,holland_apj_889_2020}.

Once the shock wave further expands, its propagation is constrained by the 
distribution of circumstellar material shaped by the supernova progenitor, 
see e.g. the case of the historical core-collapse supernova remnant 
Kes~73~\citep{katsuda_apj_863_2018}. 
The reflection towards the center of the explosion of the shock wave 
against its own circumstellar matter produces a central region emitting 
thermal X-rays that is surrounded by a radio shell. This class of supernova 
remnants is called mixed-morphology supernova 
remnant~\citep{yusefzadeh_apj_585_2003,arias_aa_622_2019,2019arXiv190908947C,
2019arXiv190906131D}. 
Most evolved supernova remnants exhibit morphologies deviating from 
sphericity. These asymmetries might have many origins, such as the intrinsic 
asymmetry of the explosion~\citep{2020A&A...636A..22O}, but also the aspherical distribution 
of circumstellar material like an asymmetric~\citep{blondin_apj_472_1996} 
wind or a  bow shocks generated by runaway massive stars. 
The organised magnetic field of the ISM imposes stellar wind bubbles of 
massive stars an elongated morphology~\citep{vanmarle_584_aa_2015}, which, 
in its turn, imposes rectangular shapes to the supernova 
remnants~\citep{meyer_515_mnras_2022}. 
On the long term, they can reflect in the subsequent pulsar wind nebulae 
of plerionic core-collapse supernova 
remnants~\citep{meyer_mnras_515_2022,bandiera_mnras_165B_2023}.

In this study, we explore by means of 2.5D MHD simulations the morphology and 
emission properties of core-collapse supernova remnants generated by runaway 
massive stars, including the rotation and magnetisation of the stellar winds. 
Furthermore, our approach made of high-resolution simulations using a uniform 
cylindrical grid~\citep{meyer_mnras_450_2015,meyer_mnras_493_2020,meyer_mnras_502_2021} 
is updated with the inclusion of a system of passive scalar tracers which permit us 
to time-dependently follow the materials successfully expelled from the stellar 
surface throughout the many evolutionary phases undergone by the high-mass star. 
The circumstellar medium of the supernova progenitor is first simulated from the 
zero-age main-sequence of the star until the supernova time using methods 
similar to that of~\citet{comeron_aa_338_1998,vanmarle_aa_444_2005,vanmarle_aa_469_2007, 
vanmarle_apj_734_2011,mackey_apjlett_751_2012,vanmarle_aa_561_2014,meyer_2014bb,  
meyer_obs_2016,meyer_mnras_464_2017,meyer_mnras_496_2020,green_aa_625_2019}. 
Then, the early supernova-wind interaction is calculated in a one-dimensional  
fashion~\citep{chevalier_apj_258_1982,truelove_apjs_120_1999,dwarkadas_apj_630_2005,
dwarkadas_apj_667_2007,whalen_apj_682_2008,vanmarle_mnras_407_2010,telezhinsky_aph_35_2012,
telezhinsky_aa_541_2012,telezhinsky_aa_552_2013,chiotellis_mnras_435_2013,
broersen_mnras_441_2014} and mapped onto 
the circumstellar medium in order to model the evolution of the supernova remnant 
up to a time $\sim 100\, \rm kyr$ after the explosion, see 
also~\citet{tenoriotagle_mnras_244_1990,tenoriotagle_mnras_251_1991, 
velazquez_apj_649_2006,vanveelen_aa_50_2009,vigh_apj_727_2011,chiotellis_aa_537_2012, 
toledo_mnras_442_2014,vanmarle_584_aa_2015,fang_mnras_464_2017,chiotellis_galax_8_2020, 
chiotellis_mnras_502_2021}. 
By performing a resolution study, we discuss the internal organisation of the supernova  
remnants, such as the mixing of material at work therein, and, by radiative transfer 
calculation, we predict their non-thermal appearance by synchrotron emission.

This paper is organised as follows. First, the numerical methods used to model 
magnetised core-collapse supernova remnants of runaway, rotating, magnetised 
Wolf-Rayet-evolving massive stars are presented in Section~\ref{sect:method}. 
We detail the internal properties of the supernova remnants of a runaway 
$35\, \rm M_{\odot}$ progenitor in Section~\ref{sect:results}. 
We further discuss our results in Section~\ref{sect:discussion} and finally draw 
our conclusions in Section~\ref{sect:conclusion}.


\section{Method}
\label{sect:method}

In this section we present the methods used to simulate the core-collapse supernova 
remnants of a runaway rotating magnetised $35\, \rm M_{\odot}$ massive star.

\subsection{Governing equations}
\label{sect:equations}

Our modelling of core-collapse supernova remnants is conducted within the frame 
of non-ideal magneto-hydrodynamics~\citep{meyer_mnras_496_2020}. The time evolution 
of a plasma volume element of the system is described by the mass continuity equation, 
\begin{equation}
	   \frac{\partial \rho}{\partial t}  + 
	   \bmath{\nabla}  \cdot \big(\rho\bmath{v}) =   0,
\label{eq:mhdeq_1}
\end{equation}
the equation for the conservation of linear momentum, 
\begin{equation}
	   \frac{\partial \bmath{m} }{\partial t}  + 
           \bmath{\nabla} \cdot \Big( \bmath{m} \textcolor{black}{\otimes} \bmath{v}  
           \textcolor{black}{-} \bmath{B} \textcolor{black}{\otimes} \bmath{B} + \bmath{\hat I}p_{\rm t} \Big)  
            =   \bmath{0},
\label{eq:mhdeq_2}
\end{equation}
the total energy conservation equation, 
\begin{equation}
	  \frac{\partial E }{\partial t}   + 
	  \bmath{\nabla} \cdot \Big( (E+p_{\rm t})\bmath{v}-\bmath{B}(\bmath{v}\cdot\bmath{B}) \Big)  
	  = \Phi(T,\rho),
\label{eq:mhdeq_3}
\end{equation}
and the equation for the evolution of the magnetic field, 
\begin{equation}
	  \frac{\partial \bmath{B} }{\partial t}   + 
	  \bmath{\nabla} \cdot \Big( \bmath{v}  \textcolor{black}{\otimes} \bmath{B} - \bmath{B} \textcolor{black}{\otimes} \bmath{v} \Big)  =
	  \bmath{0},
\label{eq:mhdeq_4}
\end{equation}
where $\bmath{m}=\rho\bmath{v}$ is the vector momentum, $\rho$ the gas density, $v$ the flow 
velocity, $p_{t}$ the total pressure and $\vec{B}$ the magnetic field vector, respectively. 
The total energy of the system reads, 
\begin{equation}
	E = \frac{p}{(\gamma - 1)} + \frac{ \bmath{m} \cdot \bmath{m} }{2\rho} 
	    + \frac{ \bmath{B} \cdot \bmath{B} }{2},
\label{eq:energy}
\end{equation}
and the sound speed closing the system of equations Eq.~\ref{eq:mhdeq_1}-\ref{eq:mhdeq_4} 
is defined as
\begin{equation}
	  c_{\rm s} = \sqrt{ \frac{\gamma p}{\rho} },
\label{eq:cs}
\end{equation}
with $\gamma=5/3$ the adiabatic index.

Optically-thin radiative cooling and heating 
is included into the equations using the laws presented in~\citet{meyer_2014bb}. 
They are represented by the function, 
\begin{equation}  
	  \Phi(T,\rho)  =  n_{\mathrm{H}}\Gamma(T)   
		   		 -  n^{2}_{\mathrm{H}}\Lambda(T),
\label{eq:dissipation}
\end{equation}
where 
\begin{equation}    
    n_{\mathrm{H}}= \frac{ \rho }{ \mu (1+\chi_{\rm He,Z}) m_{\mathrm{H}} }, 
\end{equation}
is the hydrogen number density, with $\mu$ the adiabatic index, $\chi_{\rm He,Z}$ 
the solar helium and metal abundance, and where the quantity 
\begin{equation}
	T =  \mu \frac{ m_{\mathrm{H}} }{ k_{\rm{B}} } \frac{p}{\rho},
\label{eq:temperature}
\end{equation}
is the temperature of the gas, respectively. The functions $\Gamma$ and $\Lambda$ 
in Eq.~\ref{eq:dissipation} stand for the heating and cooling rates for photoionised 
gas per unit time. They account for cooling from H, He and metals 
Z~\citep{wiersma_mnras_393_2009} within the solar abundance~\citep{asplund_araa_47_2009}, 
which largely dominates the cooling at high T. It additionally includes components 
for hydrogen recombination~\citep{hummer_mnras_268_1994}, collisionally excited 
forbidden line emission for the metals of atomic number $>2$ such as O, C and 
N~\citep{raga_apjs_109_1997}. The heating rate represents the photon emitted 
by the photosphere of the massive star, which ionize the recombining hydrogenoic 
ions and provoke the release of energetic electrons~\citep{osterbrock_1989}.

The equations are integrated using the {\sc pluto} 
code\footnote{http://plutocode.ph.unito.it/}~\citep{mignone_apj_170_2007,
migmone_apjs_198_2012,vaidya_apj_865_2018} using a Godunov-type solver composed of 
the shock-capturing HLL Rieman solver~\citep{hll_ref} and the eight-wave 
finite-volume algorithm~\citep{Powell1997} for magnetised flows, which makes sure 
that the magnetic field remains divergence-free, 
\begin{equation}
	\vec{ \nabla } \cdot  \vec{B} = 0,
\label{eq:divB}
\end{equation}
throughout the whole computational domain and over the entire simulation time, from the 
onset of the stellar wind to the middle-age evolution phase of the supernova remnant. 
It has been utilised in the precedent studies of the same authors devoted to the 
circumstellar medium of evolving and dying massive stars~\citep{meyer_mnras_506_2021}. 
Furthermore, the numerical scheme uses a third-order Runge-Kutta time integrator, 
the minmod flux limiter and the WENO3 interpolation scheme for the reconstruction 
of the primitive variables of the system. The time-marching algorithm is controlled 
by the Courant-Friedrich-Levy rule that is initialised to 
$C_{\rm cfl}=0.1$~\citep{meyer_mnras_507_2021}.

\begin{figure*}
        \centering
        \includegraphics[width=0.45\textwidth]{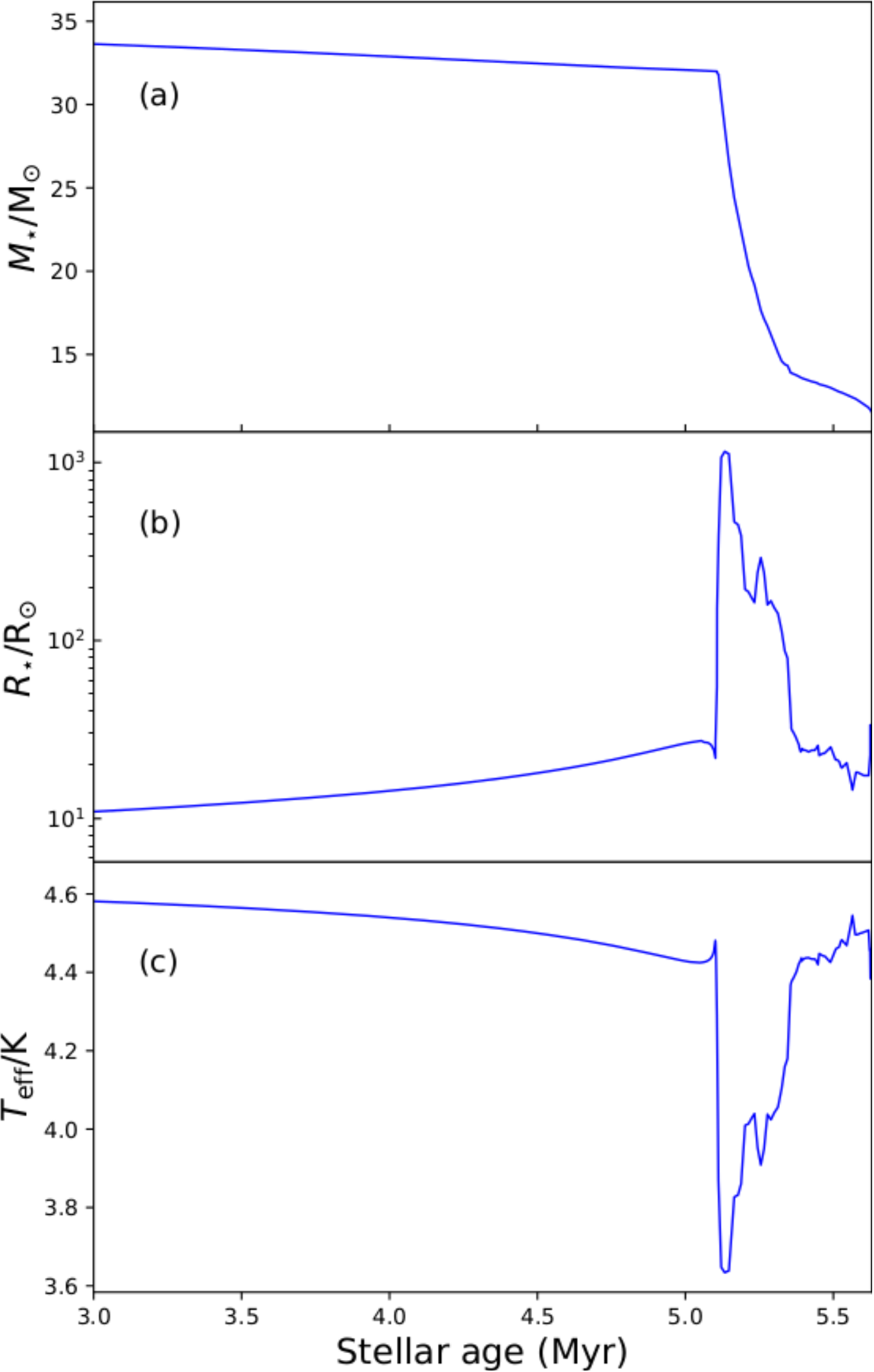}   
        \centering
        \includegraphics[width=0.46\textwidth]{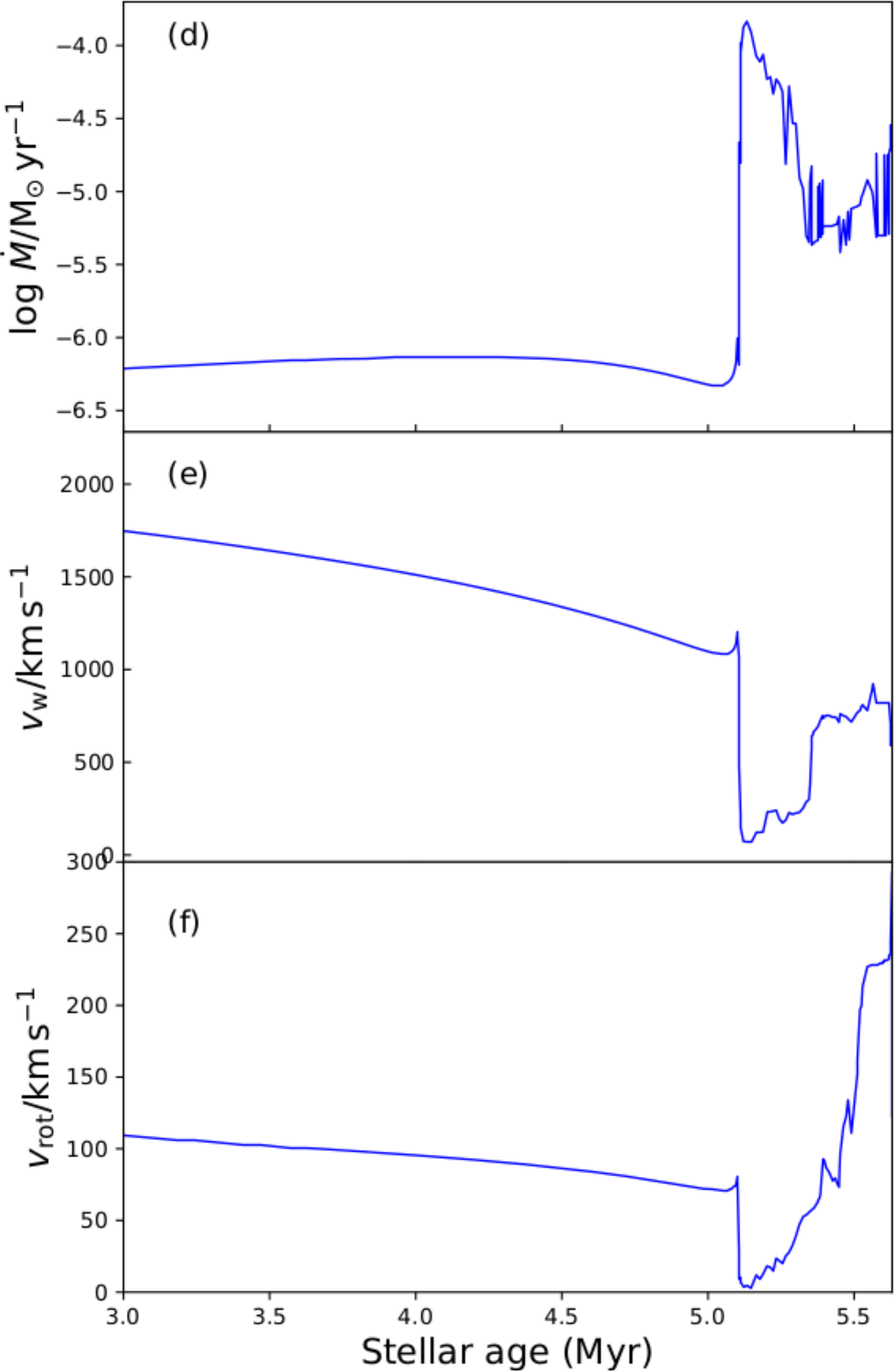}  
        \caption{
        Surface and stellar wind properties of a $35\, \rm M_{\odot}$ supernova progenitor. 
        The left-hand figures display the time evolution of the stellar mass $M_{\star}$ 
        (in $\rm M_{\odot}$, panel a), the stellar 
        radius $R_{\star}$ (in $\rm R_{\odot}$, panel b), 
        the effective temperature $T_{\rm eff}$ (in $\rm K$, panel c).
        The right-hand figures show the mass-loss rate $\dot{M}$ (in $\rm M_{\odot}\, 
        \rm yr^{-1}$, panel d), the terminal wind velocity $v_{\rm w}$ (in $\rm km\, 
        \rm s^{-1}$, panel e) and the  equatorial rotation profile of the star 
        $v_{\rm rot}$ (in $\rm km\, \rm s^{-1}$, panel f), respectively. 
        }
        \label{fig:SE_model_1}  
\end{figure*}

\subsection{Modelling the circumstellar medium}
\label{sect:method_csm}

\subsubsection{Stellar wind density}

The circumstellar medium of massive stars is modelled using a 2.5D cylindrical 
coordinate system (R,z) and a $[0;100]\times[-50;50]\, \rm pc^{2}$ computational 
domain of $8000 \times 8000$ grid zones. 
The stellar wind of the massive star is injected at the surface of a sphere of 
radius $r_{\rm in}20$ cells centered onto the origin $O (0;0;0)$. The density 
is taken to be the profile
\begin{equation}
	\rho_{w}(r=r_{\rm in},t) = \frac{ \dot{M}(t) }{ 4\pi r^{2} v_{\rm w}(t) }, 
    \label{eq:wind}
\end{equation}
where $\dot{M}$ is the mass-loss rate of the massive star, $v_{\rm w}$ the 
terminal wind velocity of the star and $r$ the radial direction. As in the previous 
papers of this series, the stellar wind nebula developing around the star is 
calculated in its reference frame. The stellar motion is mimiced by imposing 
a gas flow $v=-v_{\star}$ at the outer boundary of the computational domain 
at $z=50\, \rm pc$, with $v_{\star}=20\, \rm pc$ the stellar speed through the 
ISM, see also~\citet{comeron_aa_338_1998,green_aa_625_2019,meyer_mnras_496_2020}. 
The uniform ambient medium is in the warm phase of the ISM, with a temperature of 
$T_{\rm ISM}\approx 8000\, \rm K$ and a magnetisation of $B_{\rm ISM}=7\, \mu \, \rm G$ 
taken to be parallel to the $Oz$ symmetry axis of the computational 
domain~\citep{meyer_mnras_464_2017}.

\subsubsection{Stellar wind velocity}

The stellar wind properties of the massive star that we consider are shown in 
Fig.~\ref{fig:SE_model_1}. The figure plots the time evolution of various stellar 
global and surface quantities such as the stellar mass $M_{\star}$ (Fig.~\ref{fig:SE_model_1}a), 
the stellar radius $R_{\star}$ (Fig.~\ref{fig:SE_model_1}b), the effective temperature 
$T_{\rm eff}$ (Fig.~\ref{fig:SE_model_1}c), as well as the mass-loss rate 
(Fig.~\ref{fig:SE_model_1}d), that are directly taken from the stellar evolution 
model of the {\sc geneva} 
library\footnote{https://www.unige.ch/sciences/astro/evolution/en/database/syclist/}
~\citep{ekstroem_aa_537_2012} with an initial mass of $35\, \rm M_{\odot}$ 
and a rotation velocity of 
\begin{equation}    
    \frac{  \Omega_{\star}(t=0) }{   \Omega_{\rm K}    } = 0.1, 
\end{equation}
and with
\begin{equation}    
     \Omega_{\star}(t) = \frac{ v_{\rm rot}(t) }{ R_{\star}(t) } 
\end{equation}
the angular frequency of the star and $\Omega_{\rm K}$ the equatorial Keplerian velocity. 
The star undergoes three distinct evolutionary phase, i.e. a hot main-sequence 
with a mass-loss rate of $\dot{M} \approx 10^{-6.3}\, \rm M_{\odot}\, \rm yr^{-1}$ 
lasting at time $\approx 5.1\, \rm Myr$, a short, $\approx 0.25\, \rm Myr$-long 
cold red supergiant phase with a dense wind of $\dot{M} \approx 10^{-4.0}\, 
\rm M_{\odot}\, \rm yr^{-1}$ and a final Wolf-Rayet phase with $\dot{M} \approx 10^{-5.0}\, 
\rm M_{\odot}\, \rm yr^{-1}$. 
The star loses most of its mass during the two last phases (Fig.~\ref{fig:SE_model_1}a) 
and the supernova time is at $\approx 0.63\, \rm Myr$. 
The wind terminal velocity is calculated as
\begin{equation}    
    \frac{ v_{\rm w}(t) }{  v_{\rm esc}(t)  } = \sqrt{ \beta(T) },
\end{equation}
or equivalently, 
\begin{equation}    
     v_{\rm w}(t) = \sqrt{ \beta(T)  \frac{  2GM_{\star}(t) }{ R_{\star}(t)} },
\end{equation}
with the escape speed of the massive star $v_{\rm esc}(t)$, its stellar radius 
$R_{\star}(t)$ (Fig.~\ref{fig:SE_model_1}b), and 
\begin{equation}
        \beta_{\rm w}(T) =
        \begin{cases}
            1.0   &  \text{if } T_{\rm eff} \le 10000\, \rm K,  \\
            1.4   &  \text{if } T_{\rm eff} \le 21000\, \rm K,  \\
            2.65  &  \text{if } T_{\rm eff} >   21000\, \rm K,  \\ 
        \end{cases}
\end{equation}
a corrective factor that is function of the effective temperature~\citep{eldridge_mnras_367_2006}, 
see Fig.~\ref{fig:SE_model_1}c. 
The evolution of $v_{\rm w}$ decreases as a function of time, as the main-sequence 
wind is supersonic ($\approx 1800\, \rm km\, \rm s^{-1}$), the red supergiant wind 
velocity is very slow ($\approx 30\, \rm km\, \rm s^{-1}$) and the final Wolf-Rayet 
wind velocity increases up to ($\approx 700\, \rm km\, \rm s^{-1}$), see 
Fig.~\ref{fig:SE_model_1}e.

\begin{figure}
        \centering
        \includegraphics[width=0.45\textwidth]{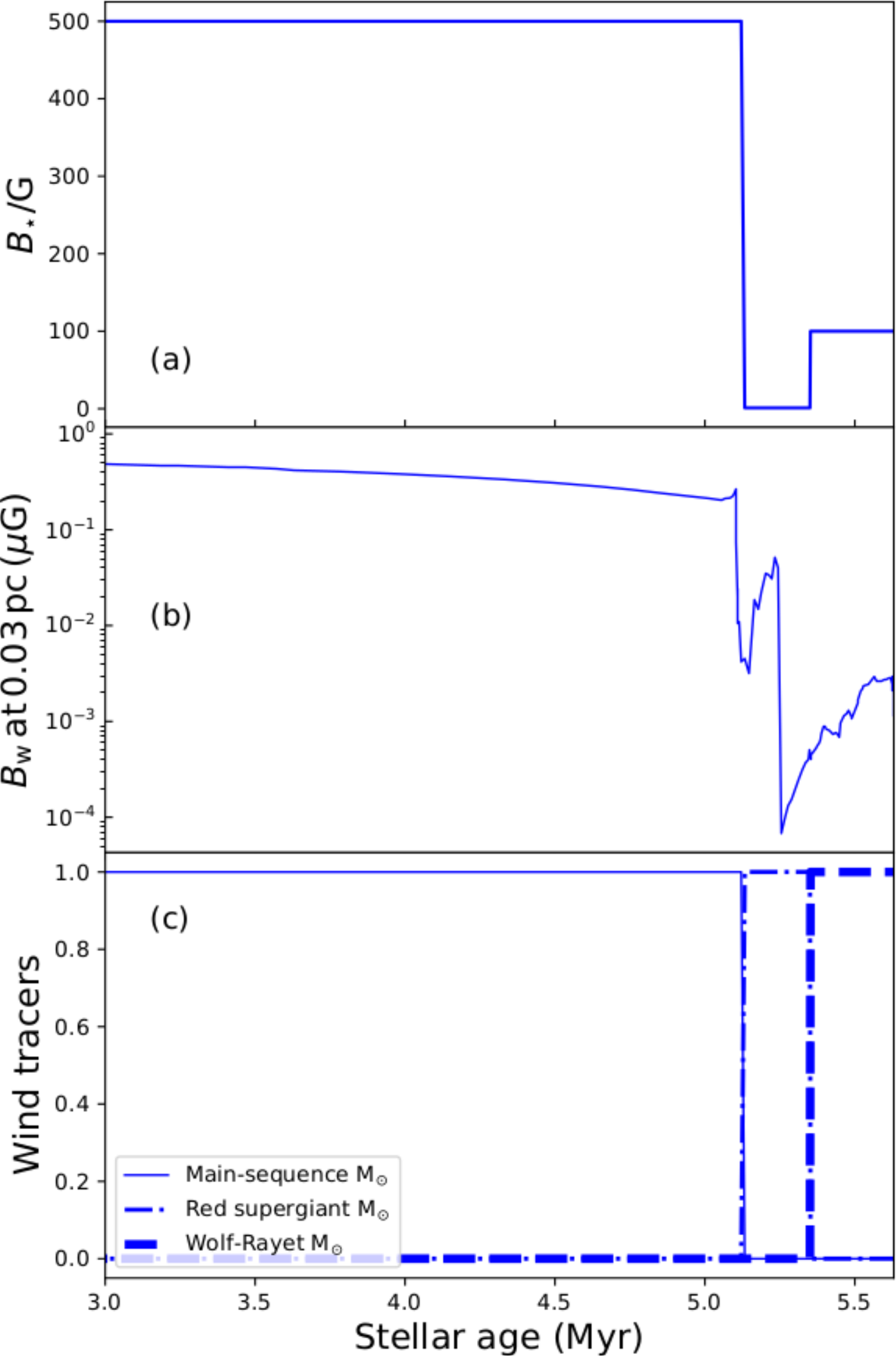}  \\
        \caption{
        \textcolor{black}{\mpo{Temporal evolution of}
        the surface magnetic field strength, $B_{\star}$,
        (panel a), the magnetic field strength in the stellar wind, 
        $B_{\rm w}$, at $r=0.03\, \rm au$ from the star (panel b),
        and the values of the passive scalar tracers (panel c) for the 
        different wind 
        phases of the stellar evolution. 
        }
        }
        \label{fig:SE_model_2}  
\end{figure}

\subsubsection{Stellar rotation and magnetic field}

The stellar magnetic field is imposed at the inner stellar wind boundary as a Parker 
spiral~\citep{parker_paj_128_1958,weber_apj_148_1967,pogolerov_aa_321_1997,pogolerov_aa_354_2000} 
that is superimposed to the stellar wind~\citep{chevalier_apj_421_1994,rozyczka_apj_469_1996,   
garciasegura_apj_860_2018,garciasegura_apj_893_2020}. It is composed 
of both a radial
\begin{equation}
	B_{\rm r}(r,t) = B_{\star}(t) \Big( \frac{R_{\star}(t)}{r} \Big)^{2},
    \label{eq:Br}
\end{equation}
and of a toroidal component
\begin{equation}
	B_{\phi}(r,t) = B_{\rm r}(r,t) 
	\Big( \frac{ v_{\phi}(\theta,t) }{ v_{\rm w}(t) } \Big) 
	\Big( \frac{ r }{ R_{\star}(t) }-1 \Big),
    \label{eq:Bphi}
\end{equation}
respectively, with 
\begin{equation}
	v_{\phi}(\theta,t) = v_{\rm rot}(t) \sin( \theta ),
\label{eq:Vphi}
\end{equation}
the latitude-corrected azimuthal component of the surface stellar velocity, 
where the factor $\sin(\theta)$ account for the latitude-dependence of the surface 
rotation (Fig.~\ref{fig:SE_model_1}f), 
The third poloidal component of the magnetic field is set to $B_{\theta}=0\, \rm G$. 
As the magnetic field is transported into the stellar wind of the massive star, 
we scale the decrease of the strength of the surface magnetic field to that of 
the Sun, via the relation
\begin{equation}    
     \frac{ B_{\star}(t) }{ B_{\odot} } 
     = \frac{ B_{\rm w}(t,r=1\, \rm au) }{ B_{\rm w,\odot}(r=1\, \rm au) },  
\end{equation}
where $B_{\rm w}(t,r=1\, \rm au)$ is the star's magnetic field at a distance of 
$r=1\, \rm au$ from the stellar center, $B_{\odot}$ the Sun's surface magnetic field 
and $B_{\rm w,\odot}(r=1\, \rm au)$ the observed strength of the magnetic field in the 
solar wind, respectively. This recipe has been used as an addition to boundary 
conditions for magnetised stellar winds in the context of massive 
stars~\citep{scherer_mnras_493_2020,herbst_apj_897_2020,baalmann_aa_634_2020,     
baalmann_aa_650_2021,meyer_mnras_506_2021}.

The time-dependence of the stellar surface magnetic field $B_{\star}(t)$ is estimated 
by assuming a typical field strength for each individual evolutionary phase of the 
massive star. The main-sequence field is taken to 
$B_{\star}=500\, \rm G$~\citep{fossati_aa_574_2015,castro_aa_581_2015,  przybilla_aa_587_2016,castro_aa_597_2017}. 
Regarding the red supergiant phase, we use the value derived for Betelgeuse, 
i.e. $B_{\star}=0.2\, \rm G$~\citep{vlemmings_aa_394_2002,vlemmings_aa_434_2005, 
kervella_aa_609_2018}. Finally, we assume the ad-hoc value of $B_{\star}=100\, \rm G$ for 
the final Wolf-Rayet phase~\citep{meyer_mnras_507_2021}. 
These assumed values are reported in Fig.~\ref{fig:SE_model_2}a and permit to build 
a magnetic field history at any radius of the stellar wind boundary $r=r_{\rm in}$, 
see Fig.~\ref{fig:SE_model_2}b.

\subsubsection{Mixing of materials}

Last, a system of passive scalar tracers permit us to time-dependently follow the 
evolution and mixing of stellar wind materials into the circumstellar medium and 
the supernova remnant. Since the $35\, \rm M_{\odot}$ of the {\sc geneva} 
library~\citep{ekstroem_aa_537_2012} undergoes three typical evolutionary 
phases, i.e. an initial long main-sequence phase, a supergiant phase and final 
Wolf-Rayet phase, we introduce the 
corresponding tracers $Q_{\rm MS}$, $Q_{\rm RSG}$ and $Q_{\rm WR}$, obeying the 
continuity equations
\begin{equation}
	\frac{\partial (\rho Q_{\rm MS}) }{\partial t } 
	+ \bmath{ \nabla } \cdot ( \bmath{v} \rho Q_{\rm MS}) = 0,
\label{eq:tracer1}
\end{equation}
\begin{equation}
	\frac{\partial (\rho Q_{\rm RSG}) }{\partial t } 
	+ \bmath{ \nabla } \cdot ( \bmath{v} \rho Q_{\rm RSG}) = 0,
\label{eq:tracer2}
\end{equation}
and
\begin{equation}
	\frac{\partial (\rho Q_{\rm WR}) }{\partial t } 
	+ \bmath{ \nabla } \cdot ( \bmath{v} \rho Q_{\rm WR}) = 0,
\label{eq:tracer3}
\end{equation}
respectively. The initial value of the passive scalar is set to $Q_{\rm i}=1$ during 
the time the star spends in the evolutionary phase $i=\{\rm MS, \rm RSG, \rm WR\}$ 
and $Q_{\rm i}=0$ otherwise. This implies that $Q_{\rm MS}+Q_{\rm RSG}+Q_{\rm WR}=1$ 
at every time of the simulation at the stellar wind boundary $r=r_{\rm in}$.

\subsection{Modelling the ejecta-wind interaction}
\label{sect:method_sn_wind}

The interaction between the supernova ejecta and the isotropic 
freely-expanding stellar wind of the progenitor is calculated using a 1D  
spherically-symmetric computational domain [$O$;$r_{\rm out}$]. 
The supernova ejecta initially occupy the $0 \le r \le r_{\rm max} < r_{\rm out}$ 
region of the domain. The corresponding density field reads,  
\begin{equation}
\rho(r) = \begin{cases}
        \rho_{\rm core}(r) & \text{if $r \le r_{\rm core}$},               \\
        \rho_{\rm max}(r)  & \text{if $r_{\rm core} < r < r_{\rm max}$},    \\
        \end{cases}
	\label{cases}
\end{equation}
where $ 0 < r_{\rm core}$ is filled by the constant high-density plateau 
\begin{equation}
   \rho_{\rm core}(r) =  \frac{1}{ 4 \pi n } \frac{ (10 E_{\rm ej}^{n-5})^{-3/2}
 }{  (3 M_{\rm ej}^{n-3})^{-5/2}  } \frac{ 1}{t_{\rm max}^{3} },
   \label{sn:density_1}
\end{equation}
while the ejecta velocity display an homologous expansion
\begin{equation}
    v(r) = \frac{ r }{ t },
\end{equation}
which is reaching 
\begin{equation}
   v_{\rm core} = \bigg(  \frac{ 10(n-5)E_{\rm ej} }{ 3(n-3)M_{\rm ej} } \bigg)^{1/2}
   \label{sn:vcore}
\end{equation}
at a distance $r=r_{\rm core}$ from the center of the explosion, respectively. 
The region $r_{\rm core} \le r \le r_{\rm max}$ contains the envelope of ejecta 
made of material of decreasing density, 
\begin{equation}
   \rho_{\rm max}(r) =  \frac{1}{ 4 \pi n } \frac{ \left(10 E_{\rm
ej}^{n-5}\right)^{(n-3)/2}  }{  \left(3 M_{\rm ej}^{n-3}\right)^{(n-5)/2}  } \frac{ 1}{t_{\rm max}^{3} } 
\bigg(\frac{r}{t_{\rm max}}\bigg)^{-n},
   \label{sn:density_2}
\end{equation}
with $n=11$, see~\citet{chevalier_apj_258_1982,truelove_apjs_120_1999}. At the edge of 
the supernova shock wave the gas flow speed is $v_{\rm max}=30000\, \rm km\, \rm s^{-1}$, 
defining the time at which we start the simulations, 
\begin{equation}
    t_{\rm max} = \frac{r_{\rm max}}{v_{\rm max}},
\end{equation}
and $r_\mathrm{max}$ is found using the numerical procedure of~\citet{vanveelen_aa_50_2009}.

In the above relations, $E_{\rm ej}=5 \times 10^{50}\, \rm erg$ is the energy released by 
the supernova explosion, and,  
\begin{equation}
   M_{\rm ej} =  M_{\star} - \int_{t_\mathrm{ZAMS}}^{t_\mathrm{SN}} \dot{M}(t)~ dt - M_{\mathrm{NS}} = 10.12\, \rm M_{\odot},
   \label{eq:co}
\end{equation}
is the ejecta mass, with $t_\mathrm{ZAMS}$ and $t_\mathrm{SN}$ the zero-age main-sequence 
and supernova times of the evolution of the $35\, \rm M_{\odot}$ that we consider, and 
$M_{\mathrm{NS}}=1.4\, \rm M_{\odot}$ the mass of the neutron star left behind the dead 
progenitor.

\subsection{Modelling the supernova remnant}
\label{sect:method_snr}

The 1D ejecta-wind interaction is mapped onto a 2.5D computational domain of size 
$[0;45]\times[-45;45]\, \rm pc^{2}$ before the supernova shock wave reaches the 
termination shock of the Wolf-Rayet stellar wind, which corresponds to a circle 
of radius $2.45\, \rm pc$ centered onto the origin. 
The mixing of the ejecta into the circumstellar medium at the pre-supernova time is 
traced by a passive scalar $Q_{\rm EJ}$ advected via, 
\begin{equation}
	\frac{\partial (\rho Q_{\rm EJ}) }{\partial t } 
	+ \bmath{ \nabla } \cdot ( \bmath{v} \rho Q_{\rm EJ}) = 0,
\label{eq:tracer4}
\end{equation}
and whose value is initialy set to $Q_{\rm EJ}=1$ in the ejecta at time $t_{\rm max}$ 
and to $Q_{\rm EJ}=0$ in the rest of the circumstellar and interstellar medium, 
respectively.

In this study, we focus on the case of a $35\, \rm M_{\odot}$ star that is static or moving with 
velocity  $v_{\star}=20$$-$$40\, \rm km\, \rm s^{-1}$ in a uniform medium of number density 
$0.79\, \rm cm^{-3}$, which implies that the supernova progenitor moves with a Mach 
number $M\approx 0$, $1$ and $2$, respectively. 
The supernova remnants are calculated with a grid resolution of $2000 \times 4000$ 
grid zones, while we perform a resolution study with the model with 
$v_{\star}=20\, \rm km\, \rm s^{-1}$. 
\textcolor{black}{
For \mpo{reference}, we compare our results with the supernova remnant model 
with $v_{\star}=0\, \rm km\, \rm s^{-1}$ of~\citet{meyer_515_mnras_2022}. 
}
We list the ensemble of performed models in Table~\ref{tab:models}.

\begin{table}
	\centering
	\caption{
	List of models and spatial resolution adopted in each simulation. 
	}
	\begin{tabular}{ll}
	\hline
	${\rm {Model}}$               &  Grid mesh    \\ 
	\hline   
	Run-35-MHD-0-SNR              &  $3000 \times 6000^{(a)}$          \\ 
	Run-35-MHD-20-SNR-1000        &  $1000 \times 2000$          \\
	Run-35-MHD-20-SNR-2000        &  $2000 \times 4000$          \\
	Run-35-MHD-20-SNR-4000        &  $4000 \times 8000$          \\
	Run-35-MHD-20-SNR-8000        &  $8000 \times 16000$          \\	
	Run-35-MHD-40-SNR-2000        &  $2000 \times 4000$          \\	
	\hline    
     (a) \citet{meyer_515_mnras_2022}
	\end{tabular}
\label{tab:models}
\end{table}


\begin{figure*}
        \centering
        \includegraphics[width=0.85\textwidth]{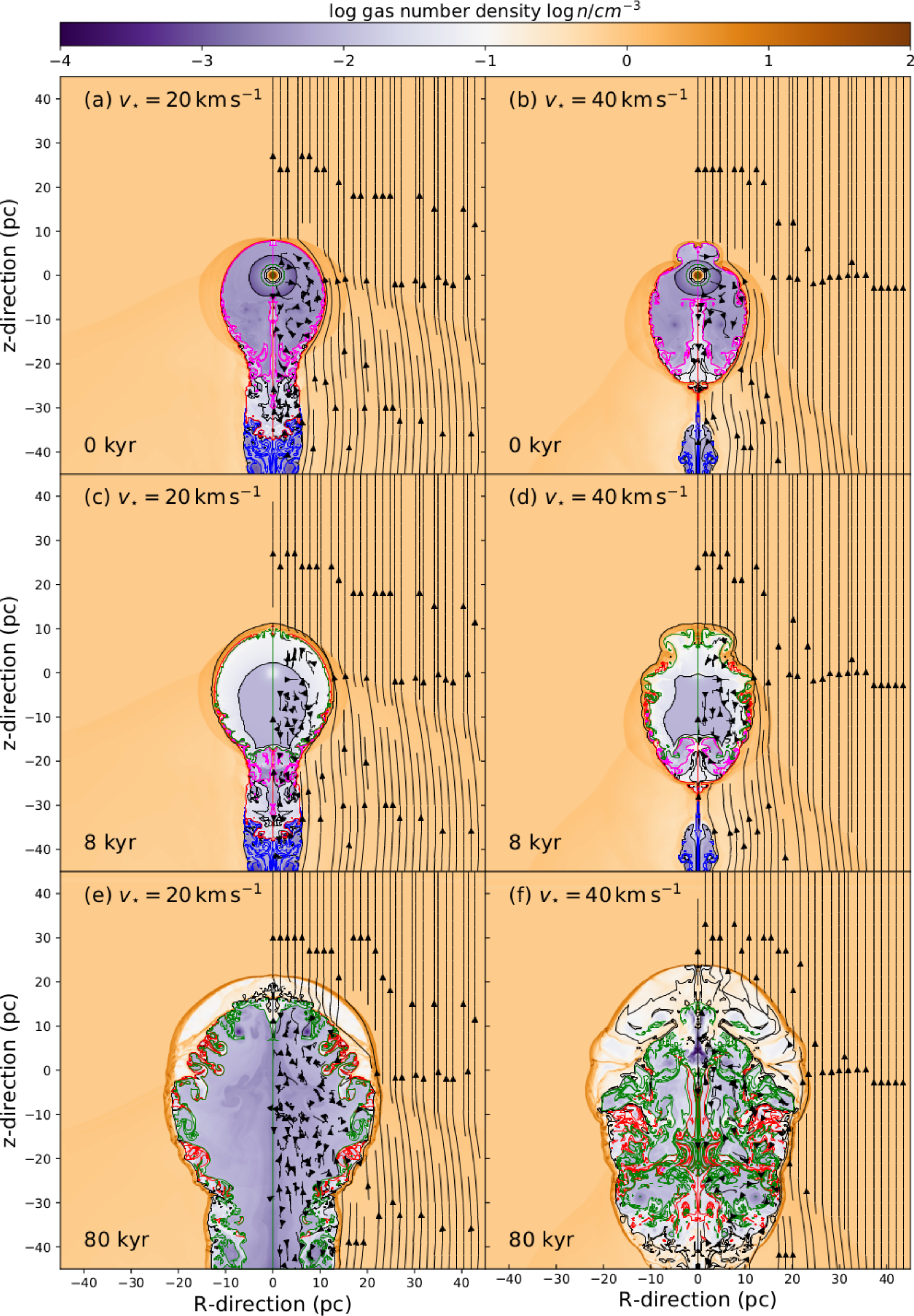}  \\
        \caption{
        Number density fields in our magneto-hydrodynamical simulation of 
        the supernova remnant of the runaway $35\, \rm M_{\odot}$ star rotating 
        with $\Omega_{\star}/\Omega_{\rm K}=0.1$ and moving with velocity 
        $v_{\star}=20\, \rm km\, \rm s^{-1}$ (left) and 
        $v_{\star}=40\, \rm km\, \rm s^{-1}$ (right). 
        The various contours highlight the region with a $50\%$ contribution of 
        supernova ejecta (green), Wolf-Rayet wind (purple), red supergiant 
        wind (red), main-sequence material (blue), respectively. 
        The black lines are iso-temperature contours ($T=10^{6}\, \rm K$) 
        and the black arrows are magnetic field lines. 
        %
        }
        \label{fig:SNR_time_evolution}  
\end{figure*}

\section{Results}
\label{sect:results}

This section presents the modelled core-collapse supernova remnant generated 
by a Wolf-Rayet-evolving runaway massive progenitor star moving at velocity 
$v_{\star}=20\, \rm km\, \rm s^{-1}$ and $v_{\star}=40\, \rm km\, \rm s^{-1}$. 
We begin to present the pre-supernova circumstellar medium and the evolution 
of the supernova remnants. We focus on the internal properties of the 
supernova remnants, their luminosities and the mixing of material happening 
therein.

\subsection{The pre-supernova circumstellar medium}
\label{sect:results_csm}

Fig.~\ref{fig:SNR_time_evolution} displays circumstellar medium at the supernova 
time in our MHD simulations of massive runaway stars moving with velocity 
$v_{\star}=20\, \rm km\, \rm s^{-1}$ (panel a) and $v_{\star}=40\, \rm km\, \rm s^{-1}$ 
(panel b). The figures represent the number density field (in $\rm cm^{-3}$) in 
the models and their right-hand part plots the ISM magnetic field lines (black arrows) 
vertical to the symmetry axis and penetrating the region of shocked material. 
The density field is structured as a large-scale bow shock generated during the main-sequence 
phase of the star's life. It has the standard features of the astropshere of moving 
stellar objects, with a bow shock facing the direction of stellar motion, separated from the star by 
a stand-off distance~\citep{baranov_sphd_15_1971}, 
\begin{equation}
	R_{\rm SO} = \sqrt{ \frac{ \dot{M} v_{\mathrm{w}} }{ 4 \pi \rho_{\mathrm{ISM}} v_{\star}^{2} } },
\label{eq:Ro}
\end{equation}
with $\rho_{\mathrm{ISM}}$ the mass density of the ISM material, 
and which overall shape can be approximated by, 
\begin{equation}
 R({\theta}) =  R_{\rm SO}  \frac{  \sqrt{ 3(1-\theta  )\mathrm{cotan}(\theta ) }  }{   \sin(\theta)    },
\label{eq:wilkin}
\end{equation}
where $\theta$ is the angle from the direction of stellar motion~\citep{wilkin_459_apj_1996}. 
A tubular low-density region shaped by the winds and a central circular cavity surrounded 
by a Wolf-Rayet shell are found at the apex of the bow shock. 
A series of contours trace the regions made of $50\%$ in number density of each kind 
of material, such as the main-sequence stellar wind (blue), the red supergiant 
(red) and the Wolf-Rayet (purple) stellar winds as well as the supernova ejecta (green).

The cone of the bow shock behind the star is more opened in the case of the 
star moving with $v_{\star}=20\, \rm km\, \rm s^{-1}$ (Fig.~\ref{fig:SNR_time_evolution}a) 
than in the case of the faster-moving star (Fig.~\ref{fig:SNR_time_evolution}b). 
The Wolf-Rayet wind, faster than the speed of stellar motion $v_{\star}$, forms a shell 
of evolved stellar wind and swept-up ISM gas when interacting bow 
shock~\citep{brighenti_mnras_273_1995,brighenti_mnras_277_1995}. 
The circumstellar medium of the faster-moving star is affected by Rayleigh-Taylor 
instabilities, both ahead of the bow shock where large eddies are seen, and in the 
tubular cavity  {in the wake of} the star, which close it and separate the main-sequence 
material from the other winds (Fig.~\ref{fig:SNR_time_evolution}b). 
The environment of the slower-moving stellar object displays a region of 
red supergiant material (red) separating the main-sequence wind (blue) from the Wolf-Rayet 
material (purple). Hence, the mixing of post-main-sequence materials is also at work 
before the supernova explosion when the star moves slowly, because of the size and unstable character of 
the low-density channel (Fig.~\ref{fig:SNR_time_evolution}a). 
Conversely, the supernova ejecta of the fast-moving star are released in a closed bubble 
of radius $\approx 10\, \rm pc$ centered onto the location of the defunct star 
(Fig.~\ref{fig:SNR_time_evolution}a), whereas the direct environment of the supernova 
of the slowly moving star is open to the trail of the bow shock 
(Fig.~\ref{fig:SNR_time_evolution}a).

\subsection{Supernova remnants}
\label{sect:results_snr}

The next panels of Fig.~\ref{fig:SNR_time_evolution} show the evolution of the 
supernova remnants at times $8\, \rm kyr$ ( middle line) and at time $80\, \rm kyr$ 
for the model with $v_{\star}=20\, \rm km\, \rm s^{-1}$ (left) and  
$v_{\star}=40\, \rm km\, \rm s^{-1}$ (right). 
They plot the reverberation of the supernova shock wave against the walls of the 
asymmetric Wolf-Rayet cavity at work in the remnants, together with the subsequent 
game of reflections and transmissions of waves happening in it. 
The imprint of the morphology of that Wolf-Rayet cavity already strongly modifies the 
geometry of the expanding forward shock at time $8\, \rm kyr$. The ejecta filled 
the region between the center of the explosion and the former termination 
shock of the pre-supernova bow shock, and the reverberated shock towards the 
center of the explosion carries the imprint of the distribution of Wolf-Rayet wind, 
see black contour for the region of gas hotter than $10^{6}\, \rm K$ in 
Fig.~\ref{fig:SNR_time_evolution}c and Fig.~\ref{fig:SNR_time_evolution}d. 
The red supergiant material (red) is distributed as a thin layer surrounding the 
ejecta region (green), and it has started to efficiently mix with the Wolf-Rayet 
material (purple) in the direction opposite of that of the progenitor's motion. 
Note that the red supergiant material distributes, because of the morphology of 
the instabilities of the pre-supernova circumstellar medium, as knots in the 
case of our faster-moving progenitor (Fig.~\ref{fig:SNR_time_evolution}d).

The bottom series of panels in Fig.~\ref{fig:SNR_time_evolution} display the 
supernova remnants at a later time of their evolution. The reverberated supernova 
shock wave propagated all through the cavity and experienced a second reflection 
against the walls of the cavity.
The overall remnant has grown in size and a top region of shocked ejecta and ISM 
gas expanding into the unperturbed ISM is clearly visible. It is more pronounced 
in the model with $v_{\star}=40\, \rm km\, \rm s^{-1}$ as a result of the smaller 
amount of mass in the Wolf-Rayet shell~\citep{meyer_mnras_450_2015}. 
The old supernova remnant is shaped as a Cygnus-loop-like nebulae in the model 
with $v_{\star}=20\, \rm km\, \rm s^{-1}$ as a result of the channelling of the 
supernova shock wave into the tunnel of low-density cavity, while that generated 
by the progenitor moving with $v_{\star}=40\, \rm km\, \rm s^{-1}$ harbors
a more ovoid morphology. 
The external envelope of the supernova remnant does not develop large scale instabilities 
as a result of the background ISM magnetic field~\citep{meyer_mnras_502_2021}.   
The Wolf-Rayet material is entirely mixed with the other kind of material, except in
some very small punctual regions (Fig.~\ref{fig:SNR_time_evolution}f), and the 
red supergiant material has efficiently mixed with the supernova ejecta. 
The main-sequence material is advected downstream the remnant and does not 
contribute to its composition.

\begin{figure*}
        \centering
        \includegraphics[width=0.8\textwidth]{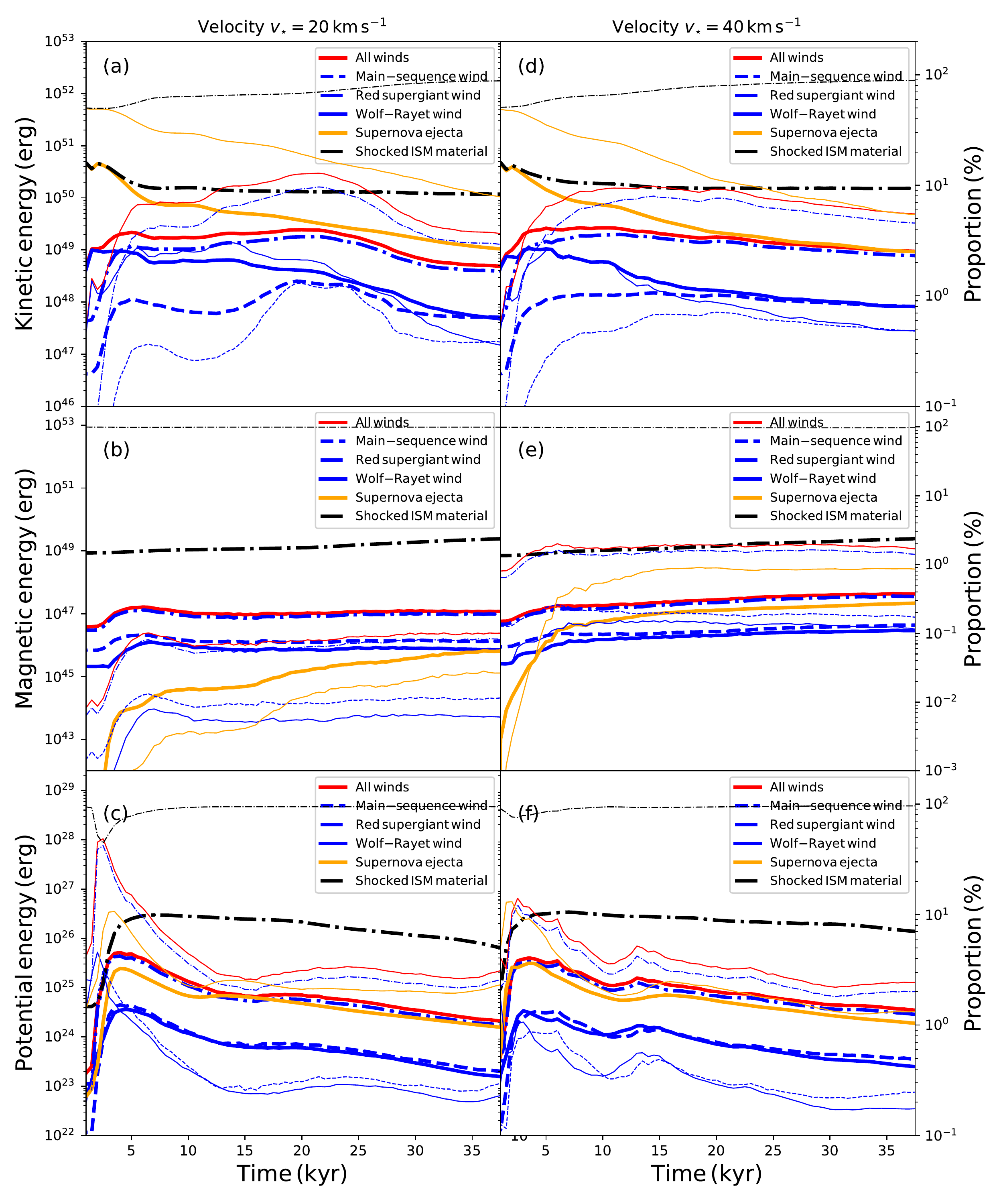} \\ 
        \caption{
        Energy evolution in the supernova remnant (thick lines) as a 
        function of time, for different simulations with 
        velocity $v_{\star}=20$ (left) and 
        $40\, \rm km\, \rm s^{-1}$ (right). 
        The figure displays the kinetic (top panel), magnetic (middle panel) 
        and potential components (bottom panel) of the total energy. 
        Each panel distinguish between the contribution from the stellar wind 
        (red solid line), main-sequence material (dashed blue line), red 
        supergiant material (dashed dotted line), Wolf-Rayet gas (solid blue 
        line), supernova ejecta (solid orange line) and shocked ISM gas (dashed 
        dotted black line). 
        The {thin lines give the corresponding percentages of the total energy at a function of time}. 
        }
        \label{fig:sn_energy}  
\end{figure*}

\subsection{Energetics}
\label{sect:results_energetics}

Fig.~\ref{fig:sn_energy} displays the kinetic (top), magnetic (middle) and 
potential energy (bottom) energies in the supernova remnants of progenitors 
moving with velocity $v_{\star}=20\, \rm km\, \rm s^{-1}$ (left) and 
$v_{\star}=40\, \rm km\, \rm s^{-1}$ (right). 
Each figure distinguishes between the components from the main-sequence (dashed blue line), 
red supergiant (dotted dashed blue line), Wolf-Rayet (solid blue line) stellar winds, 
the supernova ejecta (orange), and the ISM material (dashed dotted black line). 
The kinetic energy of a given species $i$ is calculated as, 
\begin{equation}
        E_{\rm kin}^{i}  =  \mu m_{\rm H} \iint_{\mathrm{SNR}}  n |\vec{v}|^2 Q_{i} \mathrm{dV},
        \label{eq:K}  
\end{equation}
by integrating the kinetic energy density over the entire supernova remnant, with $Q_{i}$ the
tracer associated to the considered species and $dV$ the volume element. 
Similarly, the magnetic energy of the gas is defined as, 
\begin{equation}
        E_{\rm mag}^{i}  = \iint_{\mathrm{SNR}} \frac{ |\vec{B}|^2 }{2} Q_{i} \mathrm{dV},
        \label{eq:B}  
\end{equation}
and, finally, the potential energy in the supernova remnant reads, 
\begin{equation}
        E_{\rm int}^{i} = \mu m_{\rm H} \iint_{\mathrm{SNR}}   n \epsilon Q_{i} \mathrm{dV} 
          =  \mu m_{\rm H} \iint_{\mathrm{SNR}}  n \frac{p}{(\gamma-1)} Q_{i} \mathrm{dV},
        \label{eq:I}  
\end{equation}
where $\epsilon=p/(\gamma-1)$ is the internal energy density. These quantities are evaluated 
throughout the supernova remnant simulation, up to a time $40\, \rm kyr$.

\begin{figure*}
        \centering
        \includegraphics[width=0.8\textwidth]{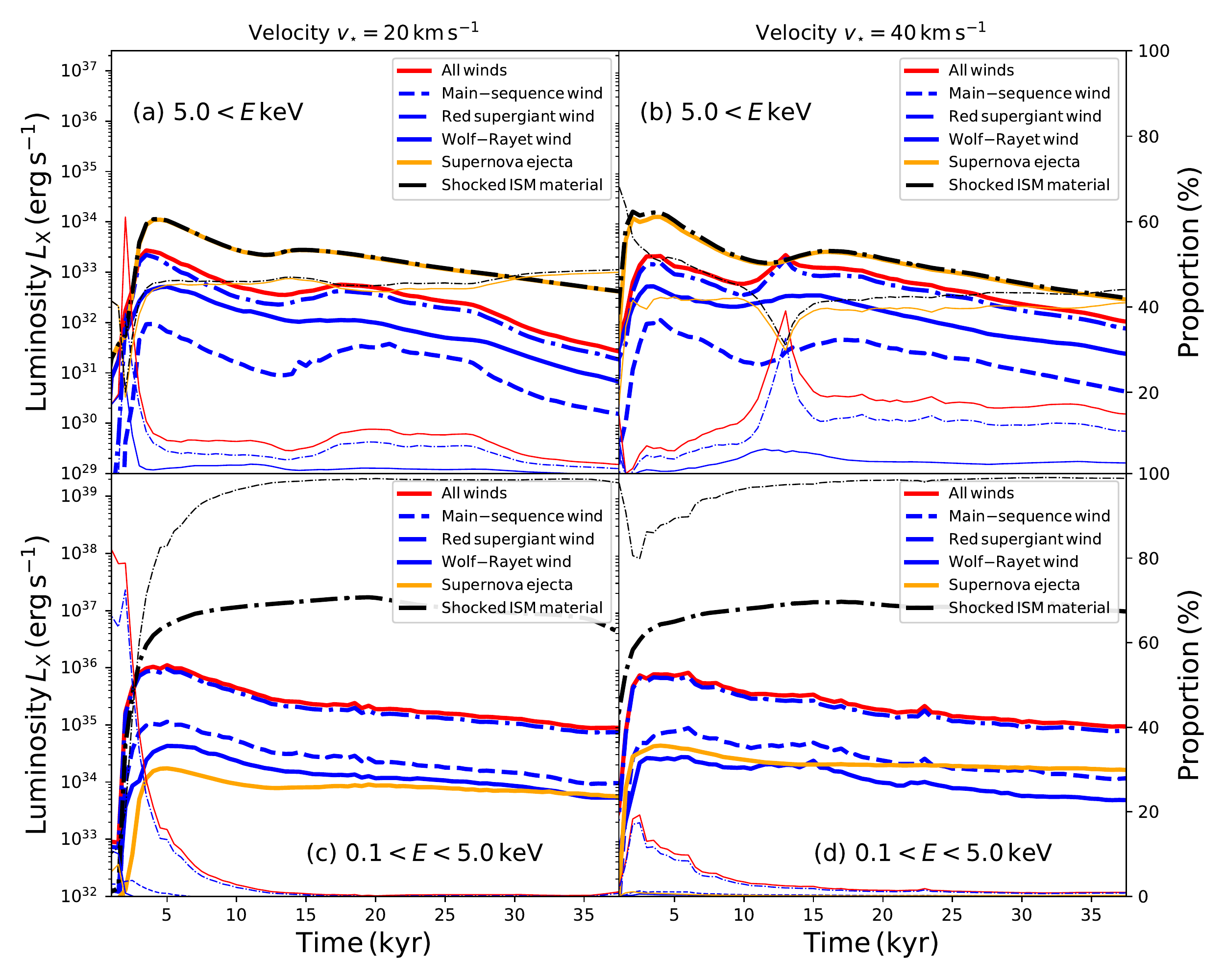}  \\        
        \caption{
        Soft X-ray  {(bottom) and hard X-ray (top}) lightcurves 
        of the supernova remnant. 
        Each panel distinguishes the contribution from the stellar wind 
        (red solid line), main-sequence material (dashed blue line), red 
        supergiant material (dashed dotted line), Wolf-Rayet gas (solid blue 
        line), supernova ejecta (solid orange line) and shocked ISM gas (dashed 
        dotted black line). 
         {We also show the time evolution of the relative contributions} to the energy budget for each 
        species in the simulation (thin lines, in per cent). 
        }
        \label{fig:sn_lum}  
\end{figure*}

In both models, the kinetic component of the energy budget of the supernova remnant 
($E_{\rm kin}\sim 10^{50}\, \rm erg$) is more important than that of the magnetic 
($E_{\rm mag}\sim 10^{49}\, \rm erg$) and internal potential ($E_{\rm pot}\sim 10^{26}\, \rm erg$) 
energy. The decreasing evolution of the ratio $E_{\rm kin}/E_{\rm mag}$  that 
in middle-age ($30$$-$$35\, \rm kyr$) core-collapse supernova remnants $E_{\rm mag}$ accounts 
for about $10\%$ of the energy of the supernova remnant and should not be neglected into 
the simulations. 
%
%
The kinetic energy of the supernova remnant whose progenitor moves with velocity 
$v_{\star}=20\, \rm km\, \rm s^{-1}$ is initially such that $E_{\rm kin}^{\rm ISM} 
\sim E_{\rm kin}^{\rm ej}$ and the stellar wind does not contribute much to the energy 
budget $E_{\rm kin}^{\rm wind} \ll E_{\rm kin}^{\rm ej}+E_{\rm kin}^{\rm ISM}$, see 
Fig.~\ref{fig:sn_energy}a. The wind contribution is governed by the red supergiant material. 
Then, the ejecta kinetic energy diminishes because of the deceleration of the blastwave 
which enters the Sedov-Taylor phase once it interacts with the circumstellar medium of 
the progenitor and sweeps up more of that material than the mass of the ejecta. 
The situation is similar in the model with progenitor's velocity $v_{\star}=40\, \rm km\, \rm s^{-1}$, 
although we note  {an energetically more important main-sequence wind in the first} $15\, \rm kyr$ 
of the remnant's life (Fig.~\ref{fig:sn_energy}d). This is due to the thinner, more unstable 
cavity of unshocked stellar wind left behind the progenitor in the case of a fast-moving 
star, which prevents the ejecta to interact directly with the main-sequence material, 
see Fig.~\ref{fig:SNR_time_evolution}.

The magnetic energy $E_{\rm mag}$ of the gas constituting the supernova remnant is 
about two orders of magnitude smaller than the kinetic energy $E_{\rm kin}$. Within the 
many materials inside of the supernova remnant, the ejecta contributes  {less than one per cent} 
to $E_{\rm mag}$, because in our models the unshocked supernova material is 
initially unmagnetised (Section~\ref{sect:method_sn_wind}). The shocked 
ejecta compress the stellar wind and the circumstellar magnetic field lines, 
which mix with the ejecta as the blastwave does through the material of the 
wind-blown bubble. 
The contribution from the stellar wind is governed by the red supergiant material, 
while the main-sequence gas, and, to an even larger extend the Wolf-Rayet material 
are negligible to the magnetic energy ($E_{\rm mag}^{\rm RSG}> E_{\rm mag}^{\rm MS}$ 
and $E_{\rm mag}^{\rm RSG}> E_{\rm mag}^{\rm WR}$). The denser surroundings generated 
by the progenitor moving with velocity $v_{\star}=40\, \rm km\, \rm s^{-1}$ induce a 
slightly more magnetised remnant (Fig.~\ref{fig:sn_energy}e). 
The internal contribution to the total energy of the supernova remnant is very 
small (Fig.~\ref{fig:sn_energy}c,f). It is regulated by the shocked ISM material 
($90\%$), the red supergiant material weights for $1\%$ of the internal energy 
while the other winds (main-sequence and Wolf-Rayet) have negligible contributions 
to $E_{\rm int}$.

\subsection{X-rays luminosities}
\label{sect:results_lum}

Fig.~\ref{fig:sn_lum} displays the hard (top) and soft (bottom) X-ray lightcurves of the 
supernova remnants of the progenitors moving with velocity 
$v_{\star}=20\, \rm km\, \rm s^{-1}$ (left) and $v_{\star}=40\, \rm km\, \rm s^{-1}$ (right). 
As in Fig.~\ref{fig:sn_energy}, the different stellar winds, the supernova ejecta, and the ISM 
material are separated using the system of passive scalar tracers. 
For a given species, the X-ray luminosity at energy $E\ge \alpha\, \rm keV$ is 
estimated as, 
\begin{equation}
        L_{X_{\textcolor{black}{\mathrm{E}\ge\alpha}}^{i}} =  \iint_{\mathrm{SNR}} j_{\mathrm{X}}^{\textcolor{black}{\mathrm{E}\ge\alpha}}(T) n_{\mathrm{H}}^{2} Q_{i} \mathrm{dV},
        \label{eq:lum_x}  
\end{equation}
where $j_{\mathrm{X}}$ is the emissivity for diffuse ISM obtained with the {\sc xspec} 
software~\citep{arnaud_aspc_101_1996}. Therefore, the luminosity in a specific energy band 
reads, 
\begin{equation}
        L_{X_{\textcolor{black}{\alpha\le\mathrm{E}\le\beta}}^{i}} =  \iint_{\mathrm{SNR}} \Big(j_{\mathrm{X}}^{\mathrm{E}\le\beta}(T)-j_{\mathrm{X}}^{\alpha\le\mathrm{E}}(T)\Big) 
        n_{\mathrm{H}}^{2} Q_{i} \mathrm{dV},
        \label{eq:lum_x_sw}  
\end{equation}
and we perform such integral for the $0.1 \le E \le 5.0\, \rm keV$ (soft) and 
$E\, \ge 5.0\, \rm keV$ (hard).

The overall hard X-ray luminosities, dominated by the ejecta and the shocked ISM, abruptly 
increase when the supernova shock wave hits the circumstellar material, before decreasing 
with time as the gas temperature lessen (thick lines of Figs.~\ref{fig:sn_lum}a,b). 
Further mild variations at time $15\, \rm kyr$ mirror the reverberation of the ejecta 
in the center of the remnants. The hard stellar-wind contribution in the model with 
$v_{\star}=20\, \rm km\, \rm s^{-1}$ accounts for $<10\%$ of the soft $L_{\rm X}$, 
while the ISM and the ejecta represent $\approx 40\%$ each. In the model with 
$v_{\star}=40\, \rm km\, \rm s^{-1}$, the wind contribution is more important, 
especially at the moment of the reflection of the shock wave inside of the remnant, 
when it can reach up to $\approx 40\%$ and exceed that of the ISM plus ejecta 
(thin lines of Figs.~\ref{fig:sn_lum}a,b). 
On the contrary, the total soft X-ray luminosity increases and reaches a plateau 
largely governed by the radiation from ISM material which  {exceeds the luminosity in hard X rays}
by several orders of magnitude. Note that the 
ejecta contribution to the soft X-ray luminosity is about $20\%$ as its temperature 
is too hot ($T\ge 10^{6}\, \rm K$) and that the stellar wind contribution is 
nebligible once the shock wave has interacted with the circumstellar medium 
(at times $5\, \rm kyr$).

Interestingly, the main-sequence wind contribution in the hard X-ray band is the smallest of all 
winds, while it is that of the Wolf-Rayet wind which is 
the least emitting in the soft X rays. In both wavebands, the red 
supergiant material dominates the stellar wind contributions to the 
total luminosity. 
This can be explained since the main-sequence wind is rapidly advected downstream 
in the ejecta-wind interaction region by the stellar motion. The Wolf-Rayet 
material efficiently mixes inside of the supernova remnant up to being confined 
in small regions scattering inside of it (Fig.~\ref{fig:SNR_time_evolution}).

\begin{figure}
        \centering
        \includegraphics[width=0.49\textwidth]{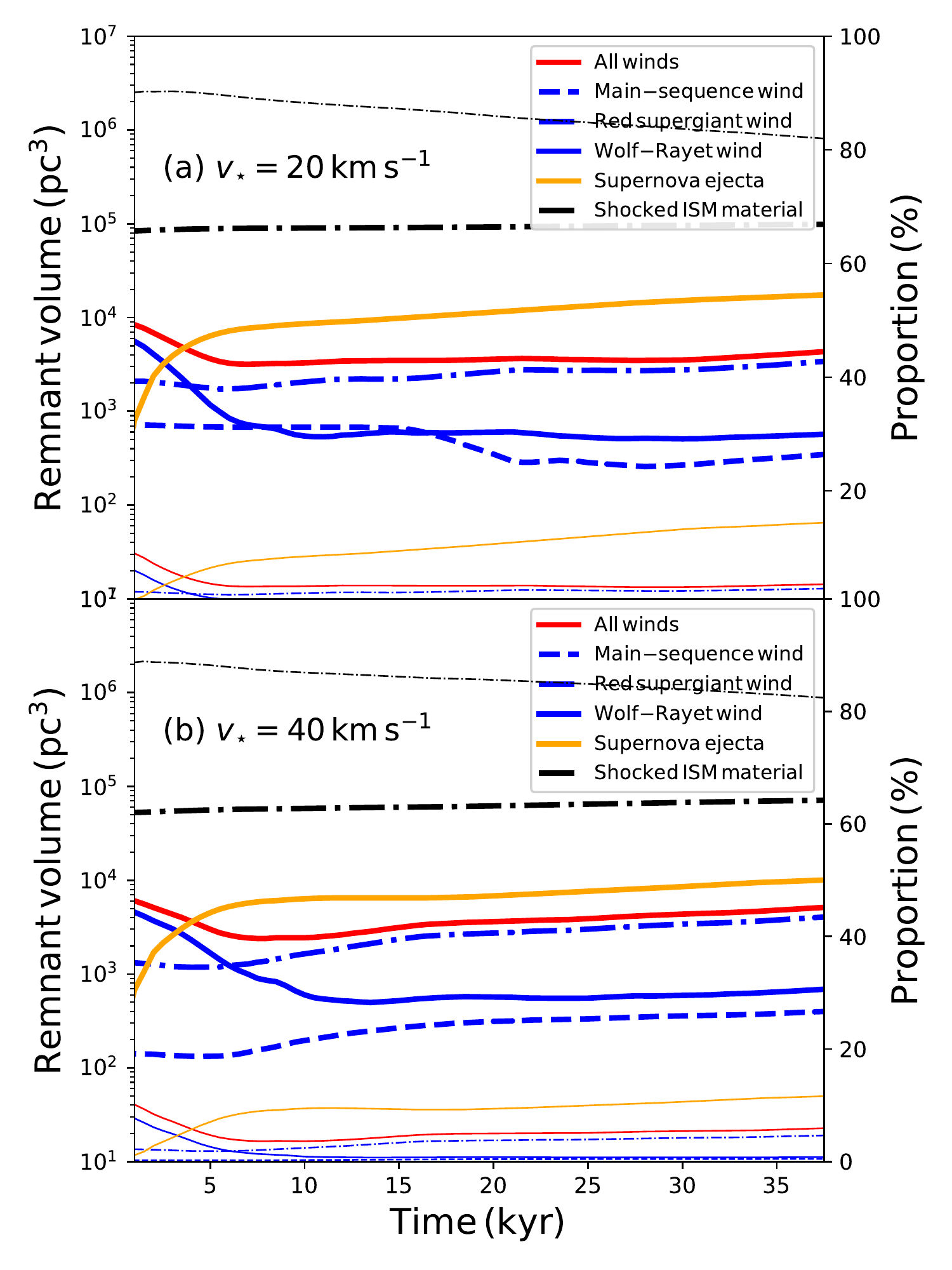}  \\       
        \caption{
        The volume (in $\rm pc^{3}$) occupied by the different 
        materials inside of the supernova remnant of the slower (top) and the faster progenitor (bottom), plotted as a function of time. 
        Each panel distinguishes between the contribution from the stellar wind 
        (red solid line), main-sequence material (dashed blue line), red 
        supergiant material (dashed dotted line), Wolf-Rayet gas (solid blue 
        line), supernova ejecta (solid orange line) and shocked ISM gas (dashed 
        dotted black line). 
        The contributions to the energy budget for each 
        species in the simulation (thin lines, in per cent) is added to the plot. 
        }
        \label{fig:sn_mass}  
\end{figure}

\subsection{Mixing of materials in the supernova remnant}
\label{sect:results_mixing}

Fig.~\ref{fig:sn_mass} plots the volume evolution of the remnants whose 
progenitors are moving with velocity $v_{\star}=20\, \rm km\, \rm s^{-1}$ (top)
and $v_{\star}=40\, \rm km\, \rm s^{-1}$ (bottom). 
Each figure distinguishes the components from the main-sequence (dashed blue line), 
red supergiant (dotted dashed blue line), Wolf-Rayet (solid blue line) stellar winds, 
the supernova ejecta (orange), and the ISM material (dashed dotted black line). 
%
%
%
%
Most of the supernova remnants are filled with shocked ISM material (see dashed 
dotted think black lines in Fig.~\ref{fig:sn_mass}a,b) which occupies about 
$80$$-$$90\%$ of its volume, all the other species representing about $10$$-$$20\%$
of the remaining volume. The volume of the remnant is constant in our simulation 
because it is calculated taking into account pre-supernova circumstellar medium, 
which does not evolve much after the time of the explosion. 
However, the volume filled by the ejecta quickly increases after the explosion during 
the freely expanding phase of the blastwave ($\le 5\, \rm kyr$ after the explosion)
and its proportion of occupied volume reaches $\approx 15\%$ in the model with 
$v_{\star}=20\, \rm km\, \rm s^{-1}$ (Fig.~\ref{fig:sn_mass}a) and $\approx 10\%$ 
$v_{\star}=40\, \rm km\, \rm s^{-1}$ (Fig.~\ref{fig:sn_mass}b) since the cavity 
of the slowly-moving progenitor is larger than that of the fast-moving one 
(Fig.~\ref{fig:SNR_time_evolution}). 
The other stellar wind materials, which very initially fill more space than the 
ejecta, quickly shrink as the blastwave propagates and mixes with the Wolf-Rayet 
stellar wind. At time $\le 10\, \rm kyr$ this induces a transition between the 
wind material in the remnant being mostly made of Wolf-Rayet gas to a 
stage in which the red supergiant gas dominates.

\begin{figure}
        \centering
        \includegraphics[width=0.47\textwidth]{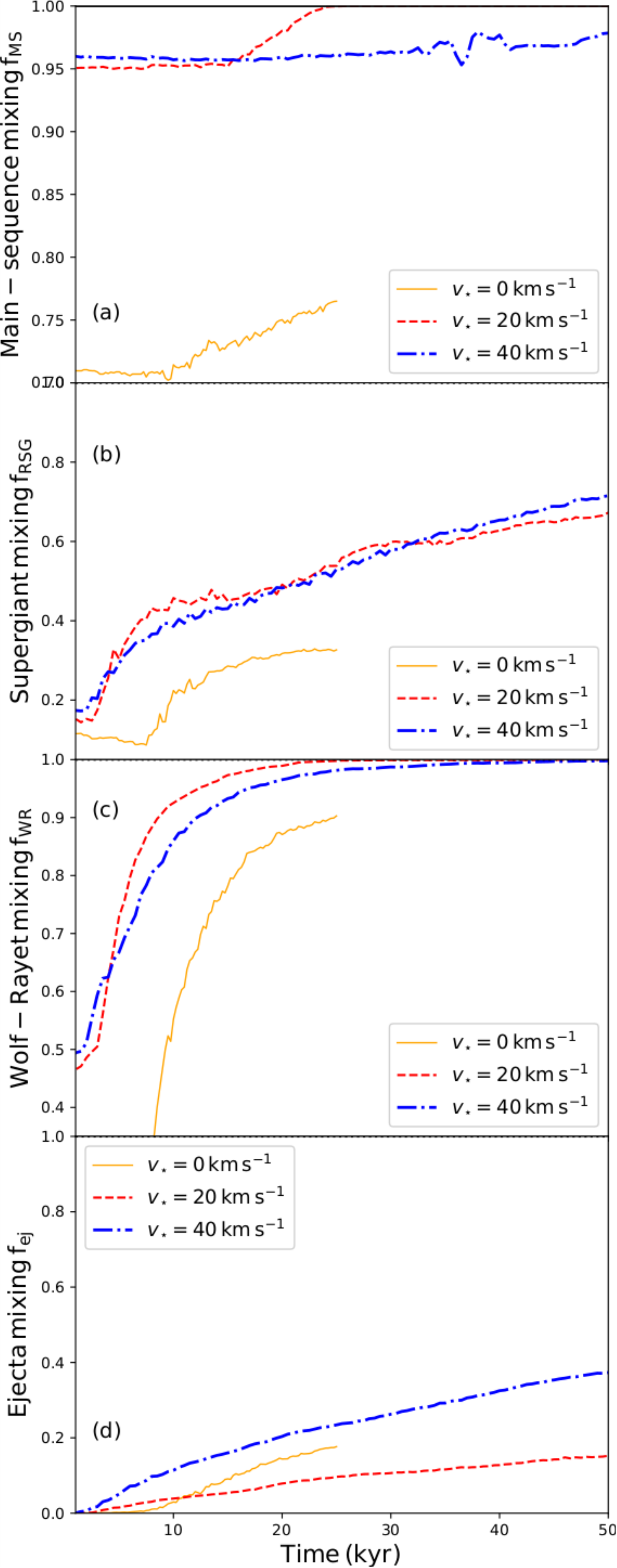}   
        \caption{
        \textcolor{black}{         
        Mixing of the different materials in the supernova remnant \mpo{during the
        first $50\, \rm kyr$ after the supernova explosion. The line colors distinguish models} with 
        stellar velocity $v_{\star}=0\, \rm km\, \rm s^{-1}$ 
        (solid orange, see~\citet{meyer_515_mnras_2022}), 
        $v_{\star}=20\, \rm km\, \rm s^{-1}$ (dotted red)
        and $v_{\star}=40\, \rm km\, \rm s^{-1}$ (dotted dashed blue). 
        The \mpo{panels are for} the 
        main-sequence (a), red supergiant (b), Wolf-Rayet (c) \mpo{wind material}
        and for the supernova ejecta (d), respectively. 
        }
        }        
        \label{fig:sn_frac}  
\end{figure}

\begin{figure*}
        \centering
        \includegraphics[width=0.7\textwidth]{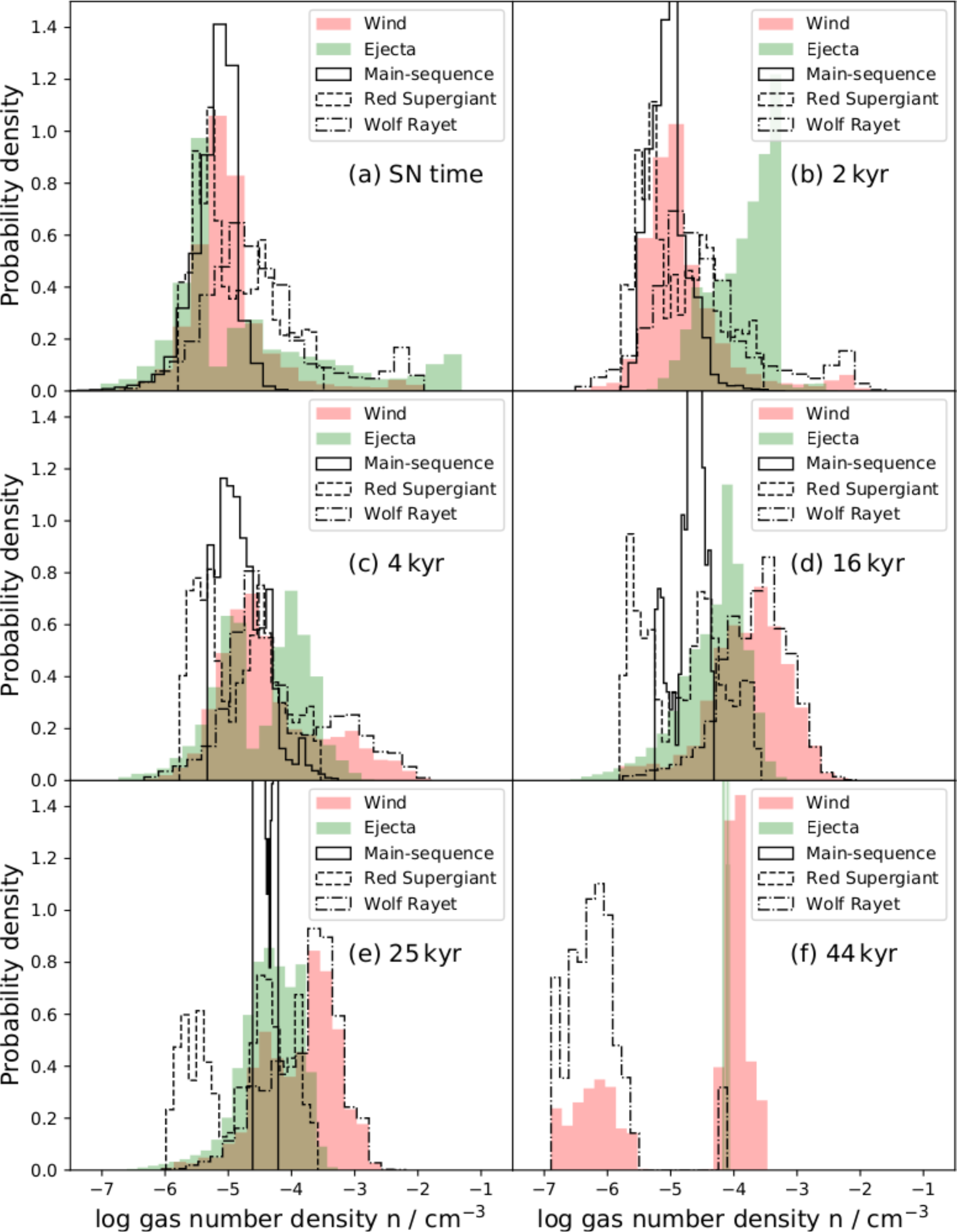}  \\        
        \caption{
        \textcolor{black}{
        Probability density distribution of the gas number density 
        (in $\rm cm^{-3}$) in the supernova remnant model with 
        $v_{\star}=20\, \rm km\, \rm s^{-1}$, plotted 
         {for various times} from the supernova explosion up to 
        $44\, \rm kyr$. 
        The figures distinguish the different kind of materials in the supernova remnants, 
        such as the stellar wind material (red) and the ejecta (green). 
        The circumstellar material of the progenitor star is further 
        divided into main-sequence (solid line), red supergiant (dotted line) 
        and Wolf-Rayet materials (dashed line), respectively. 
        }
        }
        \label{fig:distribution}  
\end{figure*}

Fig.~\ref{fig:sn_frac} displays the mixing factor for the different kind of materials present 
in the supernova remnants  {of the progenitors moving 
with $v_{\star}=20\, \rm km\, \rm s^{-1}$ (dashed red line) and with 
$v_{\star}=40\, \rm km\, \rm s^{-1}$ (dotted dashed blue line).} We define the mixing factor of the main sequence, 
\begin{equation}
        f_{\rm MS} =   \frac{ \mu m_{\rm H} \iint_{ Q_{\rm MS}\le 0.5 }  
        n Q_{\rm MS} \mathrm{dV} }{ \mu m_{\rm H} \iint_{\mathrm{SNR}}   n Q_{\rm MS} \mathrm{dV} }
          =   \frac{ \iint_{ Q_{\rm MS}\le 0.5 }  
        n Q_{\rm MS} \mathrm{dV} }{ \iint_{\mathrm{SNR}}   n Q_{\rm MS} \mathrm{dV} },
        \label{eq:f_MS}  
\end{equation}
the red supergiant, 
\begin{equation}
        f_{\rm RSG}  =   \frac{ \iint_{ Q_{\rm RSG}\le 0.5 }  
        n Q_{\rm RSG} \mathrm{dV} }{ \iint_{\mathrm{SNR}}   n Q_{\rm RSG} \mathrm{dV} },
        \label{eq:f_RSG}  
\end{equation}
and the Wolf-Rayet
\begin{equation}
        f_{\rm WR}  =   \frac{ \iint_{ Q_{\rm WR}\le 0.5 }  
        n Q_{\rm WR} \mathrm{dV} }{ \iint_{\mathrm{SNR}}   n Q_{\rm WR} \mathrm{dV} },
        \label{eq:f_WR}  
\end{equation}
stellar winds, respectively. The mixing factor of the ejecta is similarly defined, 
\begin{equation}
        f_{\rm ej} 
          =   \frac{ \iint_{ Q_{\rm ej}\le 0.5 }  
        n Q_{\rm ej} \mathrm{dV} }{ \iint_{\mathrm{SNR}}   n Q_{\rm ej} \mathrm{dV} },
        \label{eq:f}  
\end{equation}
where the numerator is the  {mass with subdominant ejecta fraction} and 
where the denominator is the initial mass of that 
same material, see~\citet{orlando_aa_444_2005}. 
For $f=0$ no mixing at all has taken place inside of the remnant while for $f=1$ 
most ejecta material is dispersed into volume elements  {to which it contributes} $<50\%$ 
in number density. 
%

\textcolor{black}{
At the beginning of the simulation, the main-sequence material is already fairly mixed to a fraction 
$\ge 95\%$, regardless of the motion of the progenitor star (Fig.~\ref{fig:sn_frac}a). 
\mpo{We consider only 
the circumstellar medium  {at distance} $<50\, \rm pc$ 
from the location of the explosion and use it} to initialise 
the simulations, which ignores the long trail of main-sequence material behind the 
star (Fig.\ref{fig:SNR_time_evolution}). 
Hence, the main-sequence material in the computational domain only represents that 
expelled by the star before the onset of the red supergiant phase. 
The mixing factor increases $15-20$  {kyr} after the explosion, when the ejecta, 
channelled into the low-density cavity, pushes the main-sequence wind out of the box, 
and $f_{\rm MS}=1$ for $v_{\star}=20\, \rm km\, \rm s^{-1}$, see Fig.~\ref{fig:sn_frac}a. 
Conversely, the red supergiant material is distributed between the Wolf-Rayet gas 
and the main-sequence bow shock, and it is compressed and mixed as the 
blastwave propagates. Since the ring produced by the Wolf-Rayet material is out 
of its stellar wind bubble, the space velocity of the progenitor star 
matters very little, and the evolution of $f_{\rm RSG}$ is similar in both cases, 
see Fig.~\ref{fig:sn_frac}b. 
}

\textcolor{black}{
The same Wolf-Rayet wind develops at early times in our supernova remnants and  
the ovoidal ring it carves is similar in both cases, and so is the circumstellar distribution 
of Wolf-Rayet material. The ring is hit by a symmetric supernova blastwave resulting from 
a spherical explosion, and the mixing factor $f_{\rm WR}$ \mpo{evolves similarly} 
in both models. 
The material is rapidly mixed, and \mpo{eventually} the Wolf-Rayet gas is distributed 
in small islands and clumps (Fig.~\ref{fig:sn_frac}c). 
The ejecta mixing factor in the supernova remnant with velocity $v_{\star}=20\, \rm km\, \rm s^{-1}$ 
stays $f_{\rm ej}<0.15$ during the early $50\, \rm kyr$ of its evolution and evolves 
little  {over the first} $20\, \rm kyr$, because once the shock wave hits the 
circumstellar medium and is slowly reflected towards the center of the location 
(Fig~\ref{fig:SNR_time_evolution}), no further mixing happens. 
In the model with velocity $v_{\star}=40\, \rm km\, \rm s^{-1}$, these reflections 
of the blastwave are facilitated by the bow shock, which permits the shock wave 
to rapidly undergo further propagation to the center of the remnant, inducing more 
mixing of the supernova ejecta with the stellar wind (Fig~\ref{fig:sn_frac}d). 
At time $30\, \rm kyr$ the mixing factor has reached $f_{\rm ej}=0.4$ to further increase 
as the waves propagates into the remnant (Fig~\ref{fig:SNR_time_evolution}f). 
All in all, within a few tens of parsecs around the center of the explosion, 
stellar motion principally affects the mixing of supernova ejecta rather than 
that of the stellar winds. 
}

\textcolor{black}{
A comparison \mpo{of} the mixing of material in supernova remnants between static and 
runaway progenitor stars is plotted as a yellow solid line in Fig.~\ref{fig:sn_frac}a-d. 
When the progenitor star does not move but evolves under the influence of the 
magnetic field of the ISM, 
the region of unshocked wind bordered by the wind/ISM contact 
discontinuity is elongated along the direction of the local ISM magnetic field, 
see \citep{meyer_515_mnras_2022}. 
The main difference \mpo{between static and runaway models} lies 
in the \mpo{paucity of} ($<70\%$) mixing of main-sequence material 
\mpo{for} $v_{\star}=0\, \rm km\, \rm s^{-1}$ (Fig.~\ref{fig:sn_frac}a). This also 
applies to the \mpo{red supergiant and Wolf-Rayet wind material}, see 
Fig.~\ref{fig:sn_frac}b,c, \mpo{whose} mixing efficiencies are much below that 
of the remnant models with runaway progenitors. 
The situation is different regarding the supernova ejecta, as its mixing efficiency 
is larger in the model with $v_{\star}=0\, \rm km\, \rm s^{-1}$ compare to that of 
with $v_{\star}=20\, \rm km\, \rm s^{-1}$ for times $>12\, \rm kyr$ (Fig.~\ref{fig:sn_frac}d). 
The reason for this is the proximity of the center of the supernova explosion with the 
shocked wind regions as an effect of the stretched contact discontinuity of stellar 
wind bubbles around static massive stars, see also  \citet{vanmarle_584_aa_2015}. 
}

\begin{figure*}
        \centering
        \includegraphics[width=0.69\textwidth]{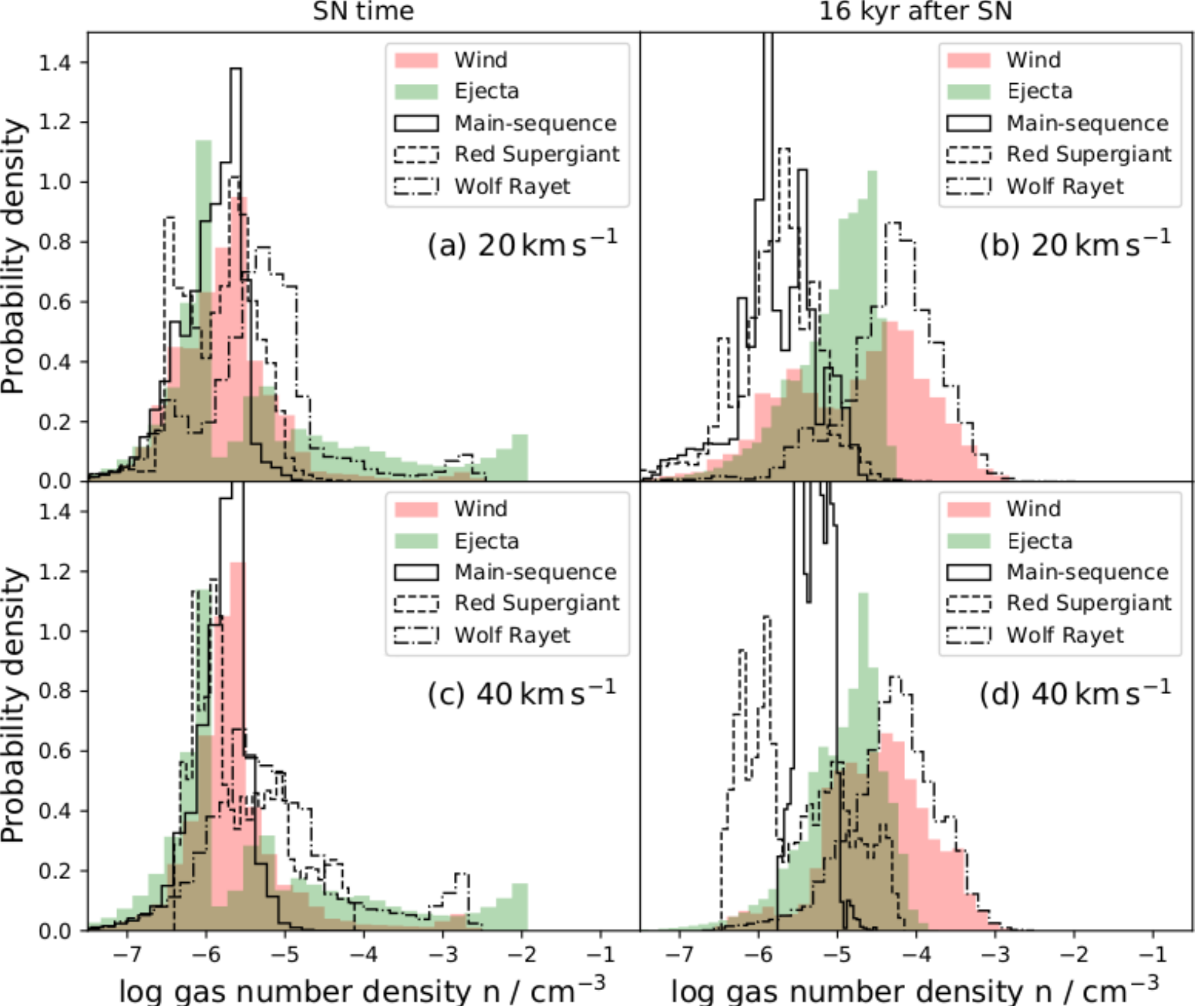}  \\        
        \caption{
        \textcolor{black}{    
        Comparison of the mixing of different materials in the supernova remnants for 
        progenitor velocity $v_{\star}=20\, \rm km\, \rm s^{-1}$ (top) and 
        $v_{\star}=40\, \rm km\, \rm s^{-1}$ (bottom)  {at the time 
        of the supernova explosion and $16\, \rm kyr$ later.} 
        The figures distinguish the stellar wind material (red) and the 
        ejecta (green). The circumstellar material of the progenitor star 
        is further divided into main-sequence (solid line), red supergiant 
        (dotted line) and Wolf-Rayet materials (dashed line), respectively. 
        }
        }
        \label{fig:distribution_comparison_rho}  
\end{figure*}

\textcolor{black}{
Fig.~\ref{fig:distribution} shows the \mpo{volumetric} distribution of the gas number density of the supernova remnant, plotted at specific times from 
the supernova explosion to the age of $44\, \rm kyr$. The 
figures distinguishe  
the stellar wind material (red) and the ejecta (green). The circumstellar material 
of the progenitor star is further divided in main-sequence (solid line), red 
supergiant (dotted line) and Wolf-Rayet (dashed line) materials, respectively. 
The \mpo{ejecta have a broad density distribution}. It is already 
interacting with the stellar wind and \mpo{displays} a two-components 
distribution, while the stellar wind distribution \mpo{has a single peak that is dominated by} main-sequence gas. 
The red supergiant gas already exhibits a rather complex distribution, 
\mpo{reflecting} the complicated and knotty morphology of the circumstellar medium 
in which the star died (Fig.~\ref{fig:distribution}a). 
At time $2\, \rm kyr$, the ejecta density distribution has evolved as a result 
of the interaction with the stellar wind and its expansion 
(Fig.~\ref{fig:distribution}b). 
At time $4\, \rm kyr$ both the stellar wind and the supernova ejecta distributions 
adopt the shape of a two-component structure. This is due to the interaction of the 
supernova shock wave with the Wolf-Rayet material, whose number density increases. 
Similarly, the red supergiant material distribution splits into two components, and 
\mpo{the two peaks become clearly separated when} the supernova ejecta are 
channelled into the low-density cavity of the circumstellar medium. As the blastwave 
further interacts with the former bow shock, the density of shocked 
Wolf-Rayet gas further increases and the distribution of stellar wind \mpo{gas shifts} to 
the higher densities (Fig.~\ref{fig:distribution}d). 
At time $25\, \rm kyr$ the red supergiant material is efficiently mixed 
a pattern 
inside of the supernova remnant, leading to multiple components in 
the red supergiant wind distribution (Fig.~\ref{fig:distribution}e), while the 
Wolf-Rayet material globally increases in density. 
At later time ($44\, \rm kyr$) only some Wolf-Rayet material remains 
unmixed, the other kinds of stellar wind being largely melted with each 
other. At this time, the supernova ejecta are almost entirely dispersed 
into the remnant. 
}

\textcolor{black}{
Fig.~\ref{fig:distribution_comparison_rho} compares the distribution of the gas 
density at selected times in the evolution of supernova 
remnants generated by a progenitor moving with velocity 
$v_{\star}=20\, \mathrm{ km\, s}^{-1}$ (top) and $v_{\star}=40\, \mathrm{ km\, s}^{-1}$ (bottom). 
The figures distinguish the stellar wind material (red) and the ejecta (green). 
The circumstellar material of the progenitor star is further divided in main-sequence, 
red supergiant and Wolf-Rayet materials, respectively. 
At the moment of the explosion, the distribution of ejecta  {does not depend on the 
progenitor speed} since we used the same explosion properties for both models 
(Fig.~\ref{fig:distribution_comparison_rho} a,c), differences being exclusively in 
the distribution of the circumstellar material and shocked ISM. The 
stellar wind material (especially red supergiant) already exhibits a different 
level of mixing because  {a faster motion of the star induces more mixing}. 
\mpo{In the model with $v_{\star}=20\, \rm km\, \rm s^{-1}$, this dichotomy increases at later times, when the red supergiant material enters the low-density cavity and the density distribution of the stellar wind and the
gas distribution} splits into two components 
(Fig.~\ref{fig:distribution_comparison_rho}b), whereas the distribution of 
Wolf-Rayet gas \mpo{appears} similar in both models (green histograms of 
Fig.~\ref{fig:distribution_comparison_rho}b,d). 
The main difference in the density distribution is the manner the main-sequence and 
red supergiant gas melt into the supernova remnant with the other kind of materials, 
see solid and dotted lined histograms of Fig.~\ref{fig:distribution_comparison_rho}b,d. 
The slightly more peaked distribution of supernova ejecta in the fast-moving 
model reflects the different reverberation properties of the supernova shock wave 
against the circumstellar medium (Fig.~\ref{fig:distribution_comparison_rho}b,d). 
}

\begin{figure}
        \centering
        \includegraphics[width=0.445\textwidth]{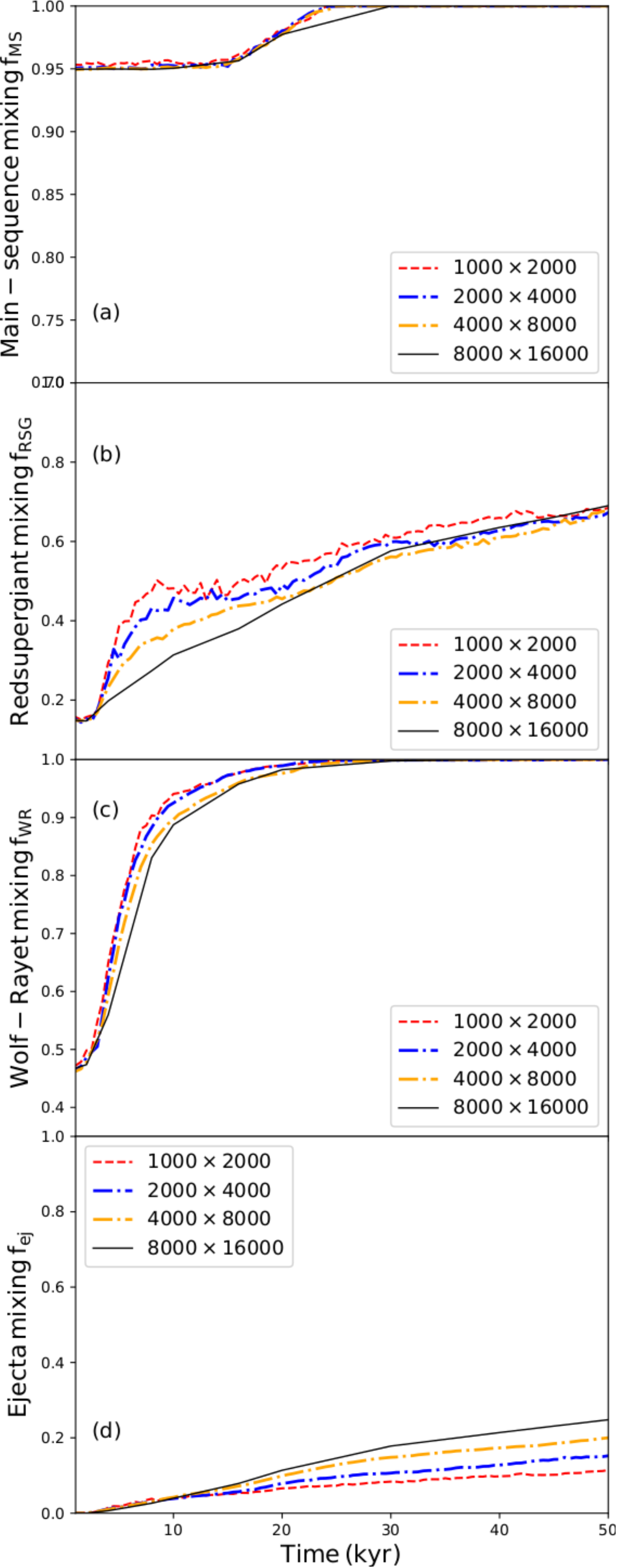} \\
        \caption{
        \textcolor{black}{ 
        Effect of spatial resolution on the 
        mixing of the different materials in the supernova remnants, 
        plotted as a function of time (in $\rm kyr$) for the models with 
        stellar velocity $v_{\star}=20\, \rm km\, \rm s^{-1}$. 
        The panels show the mixing efficiency in simulations 
        performed with spatial resolutions $1000 \times 2000$ (dashed red line),  
        $2000 \times 4000$ (dotted dashed blue line),  $4000 \times 8000$ 
        (solid orange line) and $8000 \times 16000$ (solid black line).         
        The figures plot the evolution of the mixing efficiency during the 
        early $50\, \rm kyr$ after the supernova explosion for the 
        main-sequence (a), red supergiant (b), Wolf-Rayet (c) stellar winds
        and for the supernova ejecta (d), respectively. 
        }
        }        
        \label{fig:sn_frac_res}  
\end{figure}

\begin{figure*}
        \centering
        \includegraphics[width=0.8\textwidth]{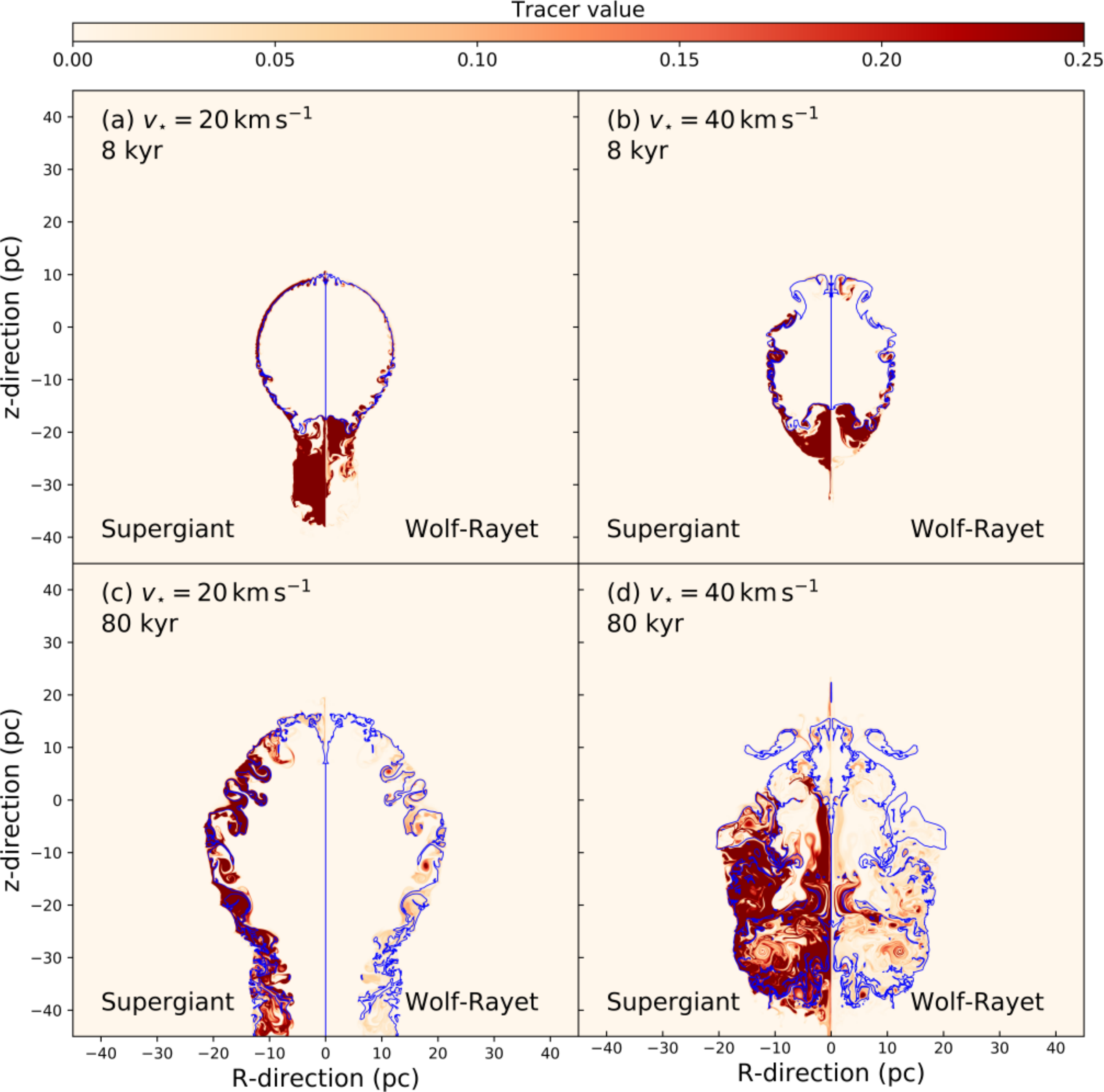}  \\        
        \caption{
        \textcolor{black}{  
        Mixing of material in the supernova remnants generated by a progenitor moving with 
        velocity $20\, \rm km\, \rm s^{-1}$ (left) and $40\, \rm km\, \rm s^{-1}$ (right), 
        shown at times $8\, \rm kyr$ (top) and $80\, \rm kyr$ (bottom). 
        The figures plot the value of the quantity $Q_{\rm RSG}$ representing the proportion 
        of red supergiant (left part of the panels) and the quantity $Q_{\rm WR}$ 
        representing the proportion of Wolf-Rayet (right part of 
        the panels) materials, respectively. 
        The blue contours of the passive scalar field $Q_{\rm EJ}$ indicate a 10 per cent 
        contribution of ejecta in number density. 
        }
        }
        \label{fig:distribution_enriched_gas}  
\end{figure*}


\section{Discussion}
\label{sect:discussion}

In this last section, we recall the limitation of the models and discuss our models in the 
context of available observations.

\subsection{Caveats of the method}
\label{sect:discussion_caveats}

Although we introduce stellar rotation and magnetised stellar wind flows in our 
simulations of core-collapse supernova remnants, our models are nevertheless 
affected by a number of limitations. 
The most important of them is the two-dimensional nature of the calculations, 
potentially impacting on the stability of the wind-ISM interaction. 
Three-dimensional simulations can solve this issue, however at much lower spatial 
resolution~\citep{katushkina_MNRAS_465_2017,katushkina_MNRAS_473_2018,   
gvaramadze_mnras_474_2018,meyer_mnras_506_2021}. 
The treatment of the ISM is also an element which can be subject to improvements, 
especially in terms of intrinsic turbulence and granulosity of the ambient medium 
in which the progenitor stars evolve~\citep{martizzi_mnras_405_2015,haid_mnras_460_2016}. 
Similarly, the interaction of the supernova shock wave with dense molecular 
regions~\citep{1975ApJ...195..715M,vandishoeck_aa_279_1993} or low-density 
parts~\citep{arias_aa_622_2019} of the ISM which might modify the propagation 
of the forward shock of the expanding supernova should also be investigated 
in future studies as this may be a key factor in the shaping of middle-age 
supernova remnants~\citep{ferreira_478_aa_2008,rathjen_mnras_504_2021}. 
Last, several physical mechanisms must be incorporated into the simulations of 
our study, such as anisotropic heat 
conduction~\citep{bedogni_190_aa_1988,velazquez_apj_601_2004, 
orlando_aa_444_2005,vieser_472_aa_2007,balsara_MNRAS_386_2008b,
balsara_MNRAS_386_2008a} or an explicit treatment of the photoionization of the 
progenitor star in order to simulate the degree of ionization of the gas in 
the remnant \citep[see e.g.][]{garciasegura_apj_517_1999,vanmarle_revmexaa_22_2004,  
arthur_apjs_165_2006,toala_apj_737_2011}.

\subsection{Effects of spatial resolution}
\label{sect:results_resolution}

Fig.~\ref{fig:sn_frac_res} repeats the analysis presented in Fig.~\ref{fig:sn_frac} 
by investigating the effect of spatial resolution on the time evolution of the mixing 
of the different materials in the supernova remnant models with stellar velocity 
$v_{\star}=20\, \rm km\, \rm s^{-1}$. 
The panels show the mixing efficiency in simulations performed with spatial resolutions 
$1000 \times 2000$ (dashed red line), $2000 \times 4000$ (dotted dashed blue line), 
$4000 \times 8000$ (dashed dotted orange line) and $8000 \times 16000$ 
(solid black line).         
The tracers in the simulation are used to separate the mixing efficiency during the early 
$50\, \rm kyr$ after the supernova explosion for the main-sequence (a), red supergiant (b), 
Wolf-Rayet (c) stellar winds and for the supernova ejecta (d), respectively. 
No notable differences in terms of mixing efficiency of the main-sequence material 
happens, especially because most of this kind of gas has left the computational domain 
at times $\ge 25\, \rm kyr$. 
\textcolor{black}{
Note that because of the definition that we adopted, the mixing factor equals to unity 
once all the material has left the computational domain (Fig.~\ref{fig:sn_frac_res}a). 
}
The mixing of the red supergiant material is the most sensitive 
to spatial resolution. The variability of the mixing factor diminishes with the increasing 
spatial resolution, and its value is slightly overestimated  when the resolution is too 
coarse (Fig.~\ref{fig:sn_frac_res}b). The models show that it has not yet converged, except at
longer times $\ge 45\, \rm kyr$, and that even higher-resolution simulations are needed 
to carefully track the mixing of the dusty supergiant material into supernova remnants. 
Inversely, the Wolf-Rayet gas mixes quickly into the remnants, and has already converged 
in our low-resolution models (Fig.~\ref{fig:sn_frac_res}c). 
The ejecta mixing behaves differently: the mixing efficiency increases by a factor of 
$\approx 2$ when the resolution is multiplied by a factor of 4 (Fig.~\ref{fig:sn_frac_res}d).
High-resolution is therefore a key ingredient of core-collapse supernova mixing 
modelling~\citep{orlando_aa_622_2019}.

\subsection{Spatial distribution of the mixed material}
\label{sect:results_location}

\textcolor{black}{
Fig. \ref{fig:distribution_enriched_gas} displays the distribution of mixed 
materials in the supernova remnant generated by a progenitor moving with velocities  
$20\, \rm km\, \rm s^{-1}$ (left panels) and $40\, \rm km\, \rm s^{-1}$ (right panels), 
shown at times $8\, \rm kyr$ (top panels) and $80\, \rm kyr$ (bottom panels). 
The figures plot the value of the quantity $Q_{\rm RSG}$ representing the proportion 
of red supergiant (left part of the panels) and the quantity $Q_{\rm WR}$ representing 
the proportion of Wolf-Rayet (right part of the panels) materials, respectively. 
The blue contours of the passive scalar field $Q_{\rm EJ}$ indicate a 10 per cent 
contribution of ejecta in number density. 
At times $8\, \rm kyr$ \mpo{there are few differences between the} models. 
The enriched gases are distributed as a ring surrounding the expanding supernova 
blastwave that is interacting with the pre-supernova circumstellar medium 
(Fig. \ref{fig:distribution_enriched_gas}a,b). The low-density cavity 
of the model with $20\, \rm km\, \rm s^{-1}$ mostly contains red supergiant  
material channelled into it (Fig. \ref{fig:distribution_enriched_gas}a).  
When the supernova remnant is much older ($80\, \rm kyr$), the manner 
the blastwave reflected towards the center of the explosion also governs \mpo{the distribution of the} stellar winds. Massive bow shocks have stopped the blastwave after 
its first reverberation \citep{meyer_mnras_450_2015,meyer_mnras_502_2021}, and both 
the red supergiant and Wolf-Rayet material are located around the 
location where the supernova shock wave interacted with the circumstellar 
medium, producing a Cygnus-loop-like region filled with ejecta and post-main-sequence 
materials (Fig. \ref{fig:distribution_enriched_gas}c). In the other model with 
$40\, \rm km\, \rm s^{-1}$, the red supergiant gas fills the lower ($z<0$) region of 
the supernova remnant, in which clumps of Wolf-Rayet-rich gas are scattered. 
This is a major difference between remnants from slow- and fast-moving progenitors, 
which has observable consequences. 
}

\subsection{Motivation of the choice of an initial $35\, \rm M_{\odot}$ progenitor} 
\label{sect:35Mo}

This work constitutes the method paper of a series of studies devoted to 
the comprehensive modelling of supernova remnants of static and runaway 
massive stars. A single, non-rotating model has already been published, 
presenting thermal and non-thermal emission maps of the transient rectangular-like 
supernova remnant generated when a supernova blastwave interacts with the elongated 
termination shock of a stellar wind bubble forming around a massive star located 
in a uniformly magnetised environment. It underlined how this mechanism may participate 
in the shaping of the rectangular supernova remnant Puppis A 
\citep{meyer_515_mnras_2022}. 
The choice of a $35\, \rm M_{\odot}$ progenitor had been motivated by 
conclusions from observations of Puppis A \citep{paron_aa_480_2008}, although 
alternative scenarios with lower-mass scenarios are currently considered
\citep{velazquez_mnras_519_2023}. Our work continues this numerical effort by 
exploring the bulk motion of the same Wolf-Rayet-evolving progenitor. 
The zero-age main-sequence rotation rate of the massive star used to simulate 
our supernova remnants is taken to $10\%$ of its break-up velocity. It is 
necessary to include rotation in the simulations otherwise the toroidal component 
of the stellar magnetic field vanishes and only the insignificantly weak radial 
component will remain. This will let the unshocked supernova ejecta unmagnetised 
and make the exercise of Fig. 4 impossible with respect to the magnetic energy. 
Further models will be performed in the future in order to continue investigating 
how the surroundings of defunct high-mass stars look like, e.g. by considering 
different initial progenitor masses and/or environments. 
A global picture regarding the mixing of enriched materials via core-collapse 
supernova remnants will hence emerge.

This collection of remnants will serve as a basis for the detailed study of 
several potential individual targets. 
Amongst others, one can mention a few supernova remnants of constrained initial 
mass consistent with a formation scenario involving a Wolf-Rayet progenitor: 
Kes 73, G292.2-0.5, G349.7+0.2, G292.0+1.8, Kes 79, Kes 17,in 
the Milky Way, N132D in the Large Magellanic Cloud or 1E0102.2-7219 in the 
Small Magellanic Cloud. 
Note that our precedent work \citep{meyer_mnras_502_2021} has established 
that supernova progenitors evolving through a Wolf-Rayet phase can produce 
bilateral and the shell-type remnants such as G296.5+10.0 and CTB109, 
respectively, at least as seen in the radio waveband by non-thermal emission. 
Such Galactic and extragalactic objects which are the perfect targets for 
tailoring our results to specific core-collapse supernova remnants. 
Last, let also note that the {\it Cherenkov Telescope Array} (CTA) will help in 
discovering new core-collapse supernova remnants to which understanding will 
require numerical models.


\section{Conclusion}
\label{sect:conclusion}

This paper investigates the evolution of young to middle-age core-collapse supernova 
remnants generated by a $35\, \rm M_{\odot}$ massive 
runaway star  {with particular emphasis on the mixing
of materials}. Our 2.5D 
magneto-hydrodynamical numerical simulations account for the 
rotation~\citep{ekstroem_aa_537_2012} and 
magnetisation~\citep{meyer_mnras_506_2021,baalmann_aa_634_2020,baalmann_aa_650_2021} of 
the progenitor star that is moving through the warm, uniform, magnetised phase of 
the interstellar medium of the Milky Way. A system of passive scalar tracers permits
us to follow the advection of the different kind of wind 
materials throughout the pre-supernova evolution of the progenitor. 
The simulations are conducted in the fashion of the precedent papers of this series, 
by first modelling the detailed circumstellar medium of the progenitor star at the 
pre-supernova time, and by releasing into it a separately calculated solution for the 
ejecta-wind that is simulated up to $100\, \rm kyr$, 
see~\citep{meyer_mnras_450_2015,meyer_mnras_493_2020,meyer_mnras_502_2021}. 
The high-resolution calculations are performed using the {\sc pluto} 
code~\citep{mignone_apj_170_2007,migmone_apjs_198_2012,vaidya_apj_865_2018} 
that has been validated for circumstellar medium and supernova remnants 
studies~\citep{meyer_2014bb,meyer_mnras_464_2017}.

It is found that the manner the stellar winds associated to the various evolutionary 
phases of the progenitor, as well as the supernova ejecta, distribute into the 
supernova remnant differs greatly depending on the bulk motion of the star, the 
level of wind mixing increasing with it, i.e. faster-moving progenitors generate older 
remnants hosting more complex, dense filamentary internal patterns of mixed species, 
where wind mixes less but ejecta does it better.  
Very little main-sequence material ($\le 5\%$) is present in the 
remnants, as it is channelled into the tubular low-density cavity of wind produced by 
the progenitor's stellar motion. The red supergiant material is the least mixed species, 
while the Wolf-Rayet gas efficiently mixes in the early $20\, \rm kyr$ after the explosion.

\textcolor{black}{
Our results show that the spatial distribution of the post-main-sequence stellar winds \mpo{depends on} the progenitor's spatial motion. At later times after the supernova explosion 
($>40\, \rm kyr$), the reflection of the supernova blastwave spreads \mpo{wind material} to the interior of the remnant nebula, distributing Wolf-Rayet clumps inside of it,  
if the progenitor's motion is important ($v_{\star}=40\, \rm km\, \rm s^{-1}$). 
The enriched stellar wind material remains in the region of the pre-supernova circumstellar medium 
for slower progenitors ($v_{\star}=20\, \rm km\, \rm s^{-1}$). 
Interestingly, the mixing efficiency of ejecta is less important in the context of 
runaway progenitors than in the case of a massive star \citep{meyer_515_mnras_2022}. 
}
The spatial resolution is  {very important} in the numerical study of mixing 
processes into core-collapse supernova remnants, as  {a low resolution} tends 
to overestimate that of the red supergiant and Wolf-Rayet stellar winds 
 {and underestimates that of the supernova ejecta}. Indeed, the regions 
filled by the winds turned to be isolate islands located around the shell made 
of the circumstellar medium shocked by the blastwave, where the material is both 
subsonic and superalfvenic.  
Such work should be extended to the stellar clusters made of multiple massive 
stars \citep{badmaev_mnras_517_2022}.


\section*{Acknowledgements}

The authors acknowledge the North-German Supercomputing Alliance (HLRN) for providing HPC 
resources that have contributed to the research results reported in this paper. 
M.~Petrov acknowledges the Max Planck Computing and Data Facility (MPCDF) for providing data 
storage resources and HPC resources which contributed to test and optimise the {\sc pluto} code.

\section*{Data availability}

This research made use of the {\sc pluto} code developed at the University of Torino  
by A.~Mignone (http://plutocode.ph.unito.it/). 
The figures have been produced using the Matplotlib plotting library for the 
Python programming language (https://matplotlib.org/). 
The data underlying this article will be shared on reasonable request to the 
corresponding author.


\bibliographystyle{mnras}
\bibliography{grid} 

\bsp	
\label{lastpage}
\end{document}